\documentclass[sigconf,nonacm]{acmart}

\usepackage{graphicx}      
\usepackage{subcaption}
\graphicspath{{figures/}}  
\usepackage{booktabs}      
\usepackage{makecell}      
\usepackage{stfloats}      
\usepackage{placeins}      
\usepackage{pifont}        
\usepackage{tikz}          
\usetikzlibrary{arrows.meta, positioning, calc, decorations.pathreplacing, shapes.geometric, backgrounds}
\usepackage{xcolor}        
\definecolor{mabBlue}{HTML}{2F6FE0}    
\definecolor{mabGreen}{HTML}{2E9E5B}   
\definecolor{mabOrange}{HTML}{EC9011}  
\definecolor{mabPurple}{HTML}{6A57D0}  
\usepackage{indentfirst}   

\newcommand{\sysname}{ArchEval}
\definecolor{mabRed}{HTML}{D33A2C}                      
\newcommand{\cmark}{\textcolor{mabGreen}{\ding{51}}}    
\newcommand{\xmark}{\textcolor{mabRed}{\ding{55}}}      
\usepackage{soul}                        
\usepackage{fvextra}                      
\definecolor{tbdblue}{HTML}{CFE6FF}

\usepackage[most]{tcolorbox}
\newtcolorbox{tbdbox}{breakable, colback=tbdblue, boxrule=0pt, sharp corners,
  left=4pt, right=4pt, top=3pt, bottom=3pt, boxsep=0pt, before skip=4pt, after skip=4pt}
\definecolor{revyellow}{HTML}{FFE9A0}
\newtcolorbox{revbox}{breakable, colback=revyellow, boxrule=0pt, sharp corners,
  left=4pt, right=4pt, top=3pt, bottom=3pt, boxsep=0pt, before skip=4pt, after skip=4pt}

\setcopyright{none}
\renewcommand\footnotetextcopyrightpermission[1]{} 
\title{\sysname{} v0.5: Measuring AI Agents as Computer Architects}

\author{%
  Chenyu Wang$^{*}$\textsuperscript{1},
  Zishen Wan$^{*}$\textsuperscript{1},
  Jeffrey Ma\textsuperscript{1},
  Shvetank Prakash\textsuperscript{1},
  Zhenting Qi\textsuperscript{1},
  Haebin Do\textsuperscript{1},
  Andy Cheng\textsuperscript{1},
  Arya Tschand\textsuperscript{1},
  Jiahe Shi\textsuperscript{2},
  Yilun Du\textsuperscript{1},
  Vijay Janapa Reddi\textsuperscript{1}%
  \vspace{3pt}
}
\affiliation{%
  \institution{%
    \textsuperscript{1}\textit{Harvard University} \quad
    \textsuperscript{2}\textit{Massachusetts Institute of Technology}%
  }
  \country{USA}
}

\definecolor{draftNotice}{HTML}{9B4F16}
\makeatletter
\g@addto@macro\@mkauthors{%
  \global\setbox\mktitle@bx=\vbox{%
    \unvbox\mktitle@bx
    \begingroup
      \centering\normalfont\fontsize{10}{12}\selectfont\color{draftNotice}
      \textbf{This is a preliminary, work-in-progress version.}\\[1pt]
      For a more complete and up-to-date version, please visit
      \href{https://comp-archeval.org}{\textcolor{draftNotice}{comp-archeval.org}}.\par
    \endgroup
    \vskip8pt
  }%
}
\makeatother

\begin{document}

\begin{abstract}
Computer architecture has long used benchmarks to make progress measurable. LLM
agents create a different measurement problem: when asked to act as computer
architects, success is not merely writing code or tuning parameters. The agent
must interpret workloads, choose mechanisms, use simulators, predict performance
before feedback, satisfy hard constraints, and decide which feasible design is
worth evaluating.

This paper introduces \sysname{}, a benchmark and platform for evaluating LLM
agents on computer architecture design and optimization. It contains 20
challenges across CPU core mechanisms, system architecture, memory systems,
accelerators, and compute-in-memory, backed by eight simulators. Each challenge
is posed under three evaluation settings: L1 \emph{full harness}, with a prepared
harness and repeated simulator feedback; L2 \emph{simulator-code container},
where simulator source is available but the agent must assemble its own
experimental workflow; and L3 \emph{agent-only}, where the agent receives static
workload evidence and constraints but no runnable simulator feedback before final
submission. Each run reports baseline-normalized verifier performance and records
the full trajectory, connecting final results to workload analysis,
simulator-tool use, prediction, constraint handling, and artifact integrity.

Initial results show a sharp boundary in current agents. With L1 support, all
four evaluated agents reach or exceed the challenge baseline and improve real
architecture designs across diverse simulators. Removing support exposes
diagnosable weaknesses: many agents do not turn simulator source into useful
local experiments, and their L3 performance predictions often disagree with
verifier results. In L3, only GPT-5.5 + Codex remains above baseline, reaching
$1.21\times$ geomean baseline-normalized performance and a 65\% win rate; the
other three agents fall below baseline. Even GPT-5.5 + Codex has only a 15\%
performance-modeling pass rate. \sysname{} therefore frames today's agents as
useful optimization assistants rather than autonomous architects, and identifies
the capabilities needed next: simulator-tool use, calibrated performance
prediction, pre-feedback design judgment, and useful mechanism discovery.
\end{abstract}

\maketitle
\begingroup
\renewcommand{\thefootnote}{\fnsymbol{footnote}}
\footnotetext[1]{Equal contribution.}
\endgroup

\section{Introduction}
\label{sec:intro}

\begin{figure}[t]
    \centering
    \includegraphics[width=\linewidth]{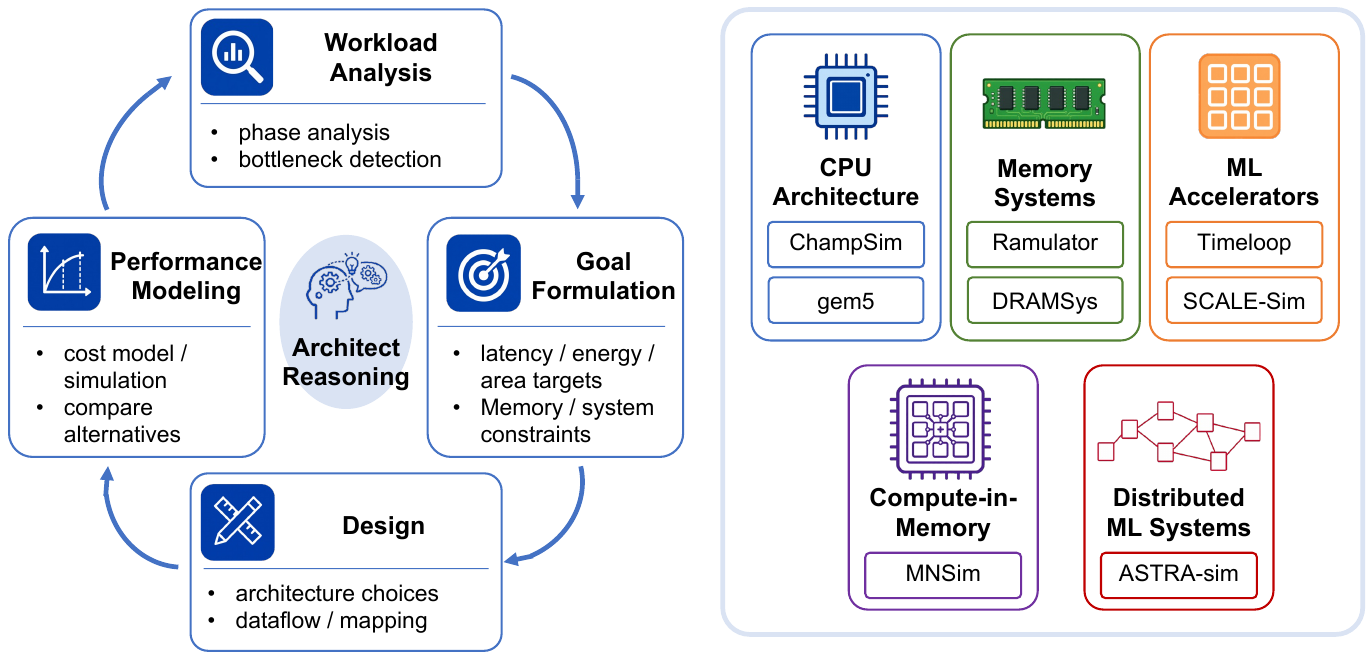}
    \caption{\textbf{\sysname{} evaluates LLM agents as computer architects.}
    (a) Existing architecture benchmarks primarily measure completed designs,
    not the design process. (b) \sysname{} asks whether LLM agents can perform
    the architecture reasoning process behind them: analyzing workloads,
    proposing mechanisms, modeling performance, and iterating across
    simulator-backed design domains.}
    \label{fig:taxonomy}
    \vspace{-0.8em}
\end{figure}

For decades, computer architecture has marked progress by building benchmarks
for the systems it learned to design: SPEC for single processors, PARSEC and
CloudSuite for workloads, and MLPerf for machine-learning hardware
\cite{spec, parsec, cloudsuite, mlperf}. These suites evaluate artifacts and
workloads. They do not evaluate the designer who chose the objective, mechanism,
simulator, and constraints. For human designers, that separation was natural:
judging the designer was more a matter of education and hiring than an
architecture-benchmark problem. That boundary is beginning to move. LLM agents
now fix real software bugs \cite{se_agents}, reason through hard problems
\cite{reasoning_agents}, solve open problems in mathematics \cite{ai_math}, and
have begun to take on computer architecture design itself
\cite{llm_dse, agentdse, archagent, agentic_architect}. Recent systems work
frames this software-to-silicon shift as a recurring challenge around tool
interfaces, verification, and feedback loops \cite{tschand2026genai}. Computer
architecture now faces a new question: \emph{can an LLM agent do the work of a computer architect, and how would we measure it?}

Consider how a computer architect actually works. Architecture design is a loop (Figure~\ref{fig:taxonomy}): analyze the workload, choose or invent a
mechanism, model how the result will perform, and refine. Within this broader
loop, architects also run optimization loops, iterating over candidate
configurations once the objective, interface, metric, and evaluator are fixed.
The classic example is design space exploration (DSE)
\cite{rl_dse, archgym, apollo}, and machine learning already automates parts of
that loop, from search to learned cost and workload models
\cite{learned_perf, perf_regression, mem_access}. The design loop is broader. It
has largely stayed human, and it is where many architectural shifts begin:
reframings like quantization, sparsity, and disaggregated serving
\cite{quantization, sparsity, disagg_serving} would not naturally emerge from
search over a narrowly predefined space. Evaluating an agent as an architect
therefore means testing both feedback-driven optimization and design-loop
judgment.

Building such a benchmark is hard for four reasons. First, there is no universal
way to compare design performance across simulators. A mathematical solution can
be checked against a known answer and a program can simply be run, but a
computer-architecture design must be measured by the simulator built for that
domain or by taping out a chip. Second, that measurement is slow: a single
simulation can take hours, days, or months, and a tape-out takes far longer.
Third, expertise is scattered across specialties, from distributed systems and
memory to CPU cores, ML accelerators, GPUs, and Compute-in-Memory (CIM). Fourth, agent autonomy is
ambiguous. A strong final result can mean that the agent designed well, but it can
also mean that humans supplied the design space, simulator workflow, and feedback
loop. The benchmark must therefore measure both the final artifact and the
support the agent needed to produce it.

\begin{figure*}[!t]
    \centering
    \includegraphics[width=0.9\textwidth]{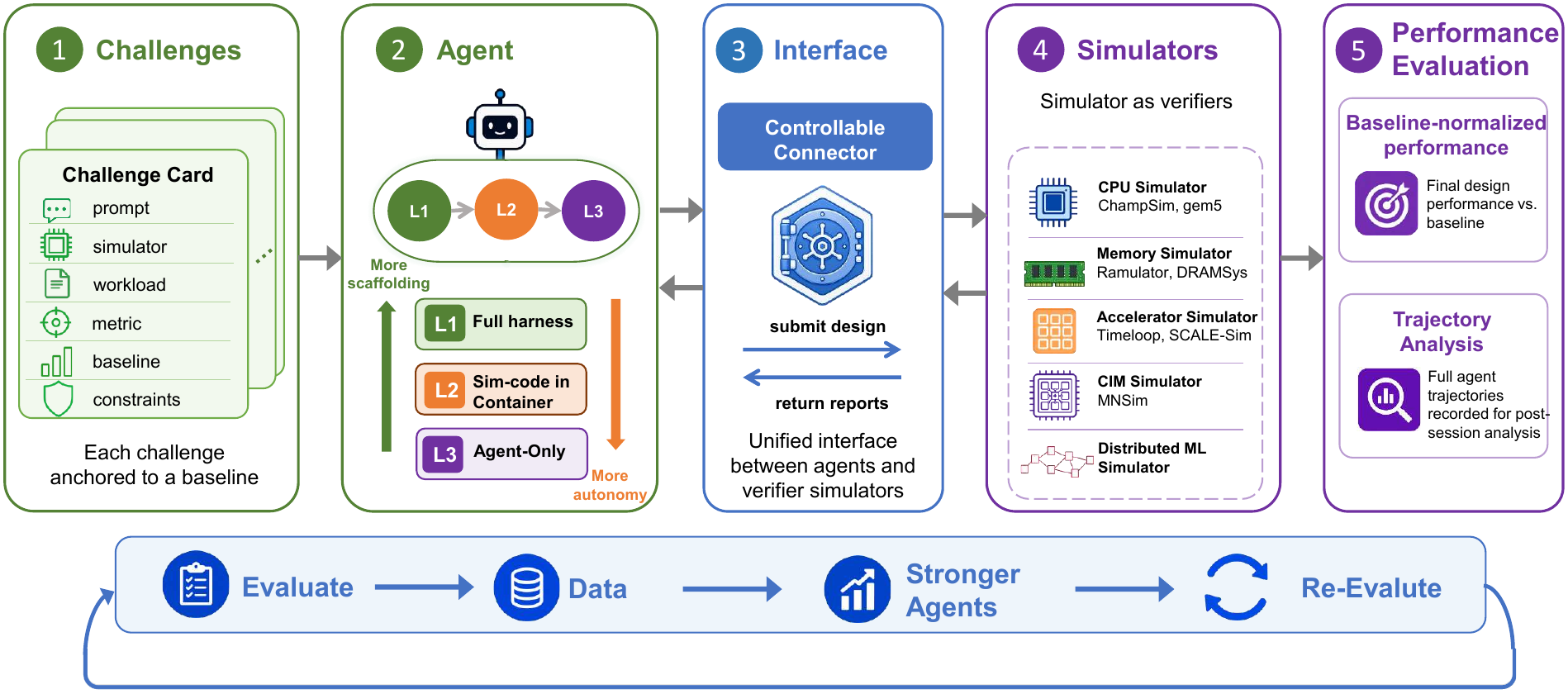}
    \caption{\textbf{\sysname{} evaluates agents through a controlled
    architecture-design workflow.} Each challenge provides a task, baseline, and
    evaluation setting; agents produce design artifacts; a connector mediates
    simulator and verifier access; and \sysname{} measures final designs while
    recording trajectories for process analysis. The resulting data can guide the
    next generation of architecture agents.}
    \label{fig:levels}
    \vspace{-0.8em}
\end{figure*}

In this work, we build \sysname{}, a systematic benchmark to evaluate LLM agents
on computer architecture design and optimization. To make comparison possible,
we put diverse simulators behind a single interface and measure every submitted
design against a credible per-challenge baseline. To make evaluation practical,
we use lightweight but simulator-backed challenges with bounded verifier
runtimes. To reflect the breadth of the field, the suite spans CPU cores, memory
systems, accelerators, distributed training, and CIM.

\sysname{}'s defining feature is that it also varies the experimental support
available to the agent. On every
challenge, the agent runs in one of three settings (Figure~\ref{fig:levels}). L1
full harness gives the agent a prepared optimization loop for design-space
exploration. L2 simulator-code container gives the simulator source and build
environment, but asks the agent to build and drive its own experimental
workflow. L3 agent-only targets the stage before such a workflow is available:
the agent receives static workload evidence and constraints, but no runnable
simulator or simulator feedback, and must estimate performance before
submission. These settings make the motivating question measurable: how much of
an architect's work can an agent already own, and which parts fail when the
prepared harness and simulator feedback are removed?

Evaluating agents across these settings gives a capability map, not just a
leaderboard. In L1, where a prepared harness supplies the experimental loop,
agents have crossed an important threshold: they can already improve real
architecture designs across diverse simulators. But the capability is uneven.
When the prepared harness is removed, performance drops in diagnosable ways. L2
shows whether an agent can use simulator source to build a working experimental
workflow. L3 shows whether it can use static workload evidence to choose a
feasible design and predict its performance before verifier-simulator feedback. In
our final single-seed suite, all four evaluated agents reach or exceed the challenge baselines at L1, but only
GPT-5.5 remains above baseline at L3; the other three agents fall below baseline
once simulator feedback is removed. The trajectory records explain the drop:
agents struggle to assemble evaluation workflows, build workload-grounded
performance models, and decide which feasible design is worth trying. The
failure is not simply artifact generation: many L3 submissions are runnable and
valid. The missing capability is design judgment before feedback, where agents
often cannot tell which feasible design is actually better. These failures are
not merely negative results; they define the roadmap for the next generation of
architecture agents.

This paper therefore makes three contributions:
\begin{itemize}
  \item \textbf{A capability map for AI computer architects.} \sysname{} frames architecture-agent evaluation as more than final design performance. It separates prediction before feedback, optimization with feedback, and generation of concrete architecture artifacts. Its L1/L2/L3 settings hold the task fixed while progressively removing the prepared harness and simulator feedback, exposing assisted DSE, simulator-tool use, and pre-feedback design judgment.
  \item \textbf{A controlled multi-simulator platform.} \sysname{} unifies eight architecture simulators behind a common connector, evaluates each design against a credible per-challenge baseline, enforces hard constraints with an isolated verifier, and records full trajectories. This makes architecture-agent evaluation simulator-grounded, comparable within each challenge, and auditable beyond a single final result.
  \item \textbf{An empirical study of current agent configurations.} We evaluate four agent configurations on 20 challenges across CPU, system, memory, accelerator, and CIM domains. The results reveal a sharp boundary: agents can improve real designs with the full harness, but often fail to build useful local experiments or judge feasible designs before simulator feedback. Trajectory analysis traces the drop to weak simulator-tool use, unreliable performance prediction, overestimated feasible designs, and limited mechanism discovery.
\end{itemize}

The remainder of this paper is organized as follows. Section~\ref{sec:framing}
develops the capability framing. Section~\ref{sec:benchmark} defines the
\sysname{} benchmark protocol. Section~\ref{sec:platform} describes the platform
that runs the protocol across heterogeneous simulators.
Section~\ref{sec:experiments} presents the empirical study by evaluation
setting, and Section~\ref{sec:capability-findings} summarizes the capability
findings that cut across those settings. Section~\ref{sec:related} positions
\sysname{} against related benchmarks and methods. Section~\ref{sec:discussion}
discusses implications, limitations, and community contributions, and
Section~\ref{sec:conclusion} concludes.

\section{Evaluating AI Agents as Architects}
\label{sec:framing}

A benchmark for AI architects should measure what an architect must do, rather
than treating one high-performing artifact as the whole answer. We organize that
work around three roles that recur in learning-enabled systems design: \emph{prediction},
\emph{optimization}, and \emph{generation}. Prediction estimates the
behavior of a candidate before the expensive measurement is available.
Optimization decides which candidate to try next when some feedback signal
exists. Generation produces the architecture artifact itself. Modern agentic
systems often combine all three, but separating them is useful because each role
fails in a different way.

Computer architecture makes this separation especially important. The final
answer is usually determined by a simulator, compiler, deployment stack, or
silicon result, but those measurements are costly and incomplete. A useful agent
must therefore predict when feedback is missing, optimize when feedback is
available, and generate concrete mechanisms under real constraints. \sysname{}
uses this capability view to ask a more precise question than whether agents can
``do architecture'': which parts of architectural work can they already own, and
which parts still require human support?

\begin{table*}[!t]
  \centering
  \scriptsize
  \renewcommand{\arraystretch}{1.12}
  \caption{\textbf{Architecture-agent capability map.}
  \sysname{} keeps the high-level roles of prediction, optimization, and
  generation, then maps each one to concrete architecture work, open capability
  questions, and the corresponding evaluation signal.}
  \label{tab:capability-map}
  \begin{tabular}{@{}p{0.07\textwidth}p{0.13\textwidth}p{0.25\textwidth}p{0.26\textwidth}p{0.21\textwidth}@{}}
    \toprule
    \textbf{Capability} & \textbf{Architecture work}
      & \textbf{What agents can plausibly attempt today}
      & \textbf{What is not yet established}
      & \textbf{How \sysname{} tests it} \\
    \midrule
    \textbf{Prediction}
      & Analyze workloads
      & Produce workload-analysis notes and inspect visible traces.
      & Whether the analysis identifies the true bottleneck and changes the
        submitted design.
      & Workload-to-design analysis. \\
      & Model performance
      & Write an agent-authored performance model or prediction script.
      & Whether the model ranks candidate designs consistently with the verifier
        simulator and calibrates uncertainty.
      & Performance-prediction diagnostic. \\
    \addlinespace[2pt]
    \textbf{Optimization}
      & Search a fixed design space
      & Edit configurations or design files, use a prepared harness, and
        improve a measured metric.
      & Whether improvement reflects broader design judgment rather than
        harness-guided search.
      & Feedback-loop optimization analysis. \\
      & Use simulator tools
      & Read source files, build tools, run scripts, and parse logs.
      & Whether the agent can assemble a reliable experimental workflow from
        simulator access.
      & Simulator-use trajectory analysis. \\
    \addlinespace[2pt]
    \textbf{Generation}
      & Choose design mechanisms
      & Recombine known policies, propose mechanisms, and tune parameters.
      & Whether the agent can connect a bottleneck to a mechanism rather than
        template-match a familiar design.
      & Design-novelty analysis. \\
      & Handle constraints
      & Follow explicit, checkable limits such as storage, area, topology,
        accuracy, and output format.
      & Whether the agent can choose the best feasible design within those
        constraints.
      & Constraint-following and shortcut/interface-failure analyses. \\
    \bottomrule
  \end{tabular}
\end{table*}

\subsection{Prediction: Modeling Architecture Behavior Before Measurement}
\label{sec:framing:prediction}

Prediction is the ability to judge a candidate design before the canonical
measurement is available. In computer architecture, that judgment may estimate
performance, energy, area, accuracy, memory traffic, communication cost, or
constraint feasibility. This is not a secondary skill. Architects routinely make
early decisions before a full simulator, compiler flow, or deployment stack can
give the final answer, let alone silicon measurement. They rely on workload
evidence, analytical models, surrogates, and experience to decide whether an idea
is worth implementing.

Learning has long served this role in systems by replacing expensive evaluation
with cheaper cost models and performance predictors
\cite{learned_perf, perf_regression, mem_access,krishnan2022automatic}. The recurring risk is the
proxy gap: a cheap predictor may correlate with the target metric on familiar
cases while misleading design under a new workload, architecture, or constraint.
For an AI architect, this gap becomes a central capability question. The agent
must know when its own model is reliable, when uncertainty is high, and when a
seemingly plausible design is likely to fail once the verifier simulator runs.

\subsection{Optimization: Searching with Evaluation Feedback}
\label{sec:framing:optimization}

Optimization is the ability to use feedback to improve a design. Once the
objective, design interface, metric, and evaluator are fixed, the task becomes a
search over candidate designs. Design space exploration (DSE) is the classic
architecture example: a human defines the design space and simulator, then a
method searches for a configuration that improves the metric
\cite{rl_dse, archgym, apollo}. This is a natural place for agents to help. A
tool-using agent can propose a candidate, read a measurement, diagnose the
failure, and try a different point.

Optimization is also where it is easiest to overstate autonomy. Strong DSE
performance does not imply that the agent has framed the architecture problem.
The design objective, legal choices, simulator path, and feedback loop may all
have been supplied by humans. DSE is therefore an important first capability, but
not the whole architect's job. The harder question is whether the agent can build
or adapt the evaluation loop when the ready-made harness is removed, and whether
it can still make sound decisions when measurement feedback is unavailable.

\subsection{Generation: Producing Architecture Artifacts and Mechanisms}
\label{sec:framing:generation}

Generation is the ability to produce the candidate architecture artifact: a
policy, mapping, configuration, microarchitectural mechanism, or co-design choice
that can be checked by architecture tools. Unlike ordinary code generation,
architecture generation has a physical interpretation. A few lines of C++ may
represent SRAM state and replacement logic; a YAML mapping may encode data
reuse, buffer pressure, and spatial parallelism; a system configuration may
determine communication distance and collective scheduling. When required, the
agent may also generate auxiliary artifacts such as design reports or
performance-model scripts, but those artifacts support prediction; the core
generated object remains the architecture design. A generated artifact is useful
only if it both satisfies the tool interface and embodies a design mechanism that
is meaningful under the workload, constraints, and target metric.

This makes generation different from text completion. The agent must translate a
high-level task into an artifact that compiles, runs, satisfies hard constraints,
and corresponds to the design argument it gives. It must also decide whether it
is merely recombining known mechanisms or proposing a substantive new one.
Generation without prediction can produce plausible but weak designs. Generation
without optimization can miss easy improvements. Generation without constraints
can win the metric by leaving the architecture problem behind.

\subsection{What an Architecture-Agent Benchmark Must Measure}
\label{sec:framing:settings}

A benchmark should make these roles observable without forcing them into a
one-to-one mapping with protocol settings. Prediction, optimization, and
generation are capability roles; evaluation settings are ways to vary the support
available to the agent. Each architecture task can involve all three roles in
different proportions. The benchmark must therefore ask what the agent predicted,
how it used feedback when feedback existed, what artifact it generated, and
whether the process that produced the artifact is visible. This is why \sysname{} poses the same challenge under multiple settings: the architecture task stays fixed, while the available evidence, tools, and feedback change.

Table~\ref{tab:capability-map} summarizes this capability map. The left column
gives the three roles, while the remaining columns make them operational as
concrete architecture work. Final performance can tell whether the submitted design
beat a baseline, but it cannot tell whether the agent predicted well, optimized
through evidence, or generated a valid architecture mechanism for the right
reason. For that reason, \sysname{} records both final performance and the
trajectory that produced it.

This framing gives four requirements for an architecture-agent benchmark. First,
it must be \emph{architecture-specific}: the agent should produce concrete architecture
artifacts rather than answer questions about architecture concepts alone. Second,
it must be \emph{execution-grounded}: designs should be measured by simulators and
baselines, because architecture quality is measured through behavior, not surface
form. Third, it must \emph{vary experimental support}, so the benchmark can distinguish an
agent that searches inside a prepared loop from one that can assemble or replace
that loop or reason before feedback exists. Fourth, it must include \emph{process
evidence}, because final baseline-normalized performance alone cannot show
whether the agent understood the workload, built a working evaluation loop,
trusted a flawed predictor, found an interface shortcut, or guessed a good
configuration. Section~\ref{sec:benchmark} turns these requirements into the
\sysname{} protocol: a challenge format, evaluation settings with different
levels of support, baseline-normalized performance, and trajectory analysis.


\section{The \sysname{} Benchmark Design}
\label{sec:method}
\label{sec:benchmark}

\sysname{} defines a simulator-grounded evaluation protocol for computer architecture
agents. Each run is specified by a challenge, an evaluation setting, an agent,
and a verifier. The challenge card in Table~\ref{tab:challenge-card} defines the
full task specification: architecture objective, workload evidence, deliverable
interface, verifier, metric, baseline, hard constraints, and visibility and
feedback policy. The setting determines what experimental support the agent can
access before submission. The verifier runs the submitted artifact against the
canonical simulator and reports validity, constraint status, native metrics, and
baseline-normalized performance.
In addition to the final performance, \sysname{} records the full trajectory so that
the submitted design can be interpreted together with the process that produced
it. This section defines that protocol, and Section~\ref{sec:platform} describes the
platform that implements it across simulators.

\subsection{What is an \sysname{} Challenge?}
\label{sec:method:format}

An \sysname{} challenge is a simulator-grounded design problem. We represent each
challenge as a \emph{challenge card}: the stable specification shared by the benchmark
and a run. Some fields are agent-visible, such as the prompt, deliverable
interface, workload evidence, metric, baseline description, and constraints.
Other fields, such as the canonical verifier script and hidden reference state,
are benchmark-internal and are mediated by the visibility and feedback policy.
The card states what the agent may see, what it must submit, what the verifier
will measure, and which constraints are non-negotiable. The
deliverable may be source code, a configuration, a mapping, a co-design artifact,
or a written model, depending on the subfield. The verifier then runs that
deliverable in the challenge's simulator and reports a typed outcome.

\begin{table}[t]
  \centering
  \footnotesize
  \setlength{\tabcolsep}{3pt}
  \renewcommand{\arraystretch}{1.15}
  \caption{\textbf{Challenge card schema.}
  Each \sysname{} challenge exposes the same protocol fields even when the
  underlying simulator and artifact type differ.}
  \label{tab:challenge-card}
  \begin{tabular}{@{}p{0.31\columnwidth}p{0.61\columnwidth}@{}}
    \toprule
    \textbf{Field} & \textbf{Meaning} \\
    \midrule
    Prompt & architecture objective and task context \\
    Deliverable & file, configuration, mapping, model, or co-design artifact to submit \\
    Workload evidence & traces, summaries, model layers, network description, or other task evidence \\
    Verifier & canonical simulator and evaluation script used by the benchmark; visibility depends on setting \\
    Metric & simulator-native metric and direction \\
    Baseline & reference design used for normalization \\
    Constraints & hard validity, resource, accuracy, topology, or format rules \\
    Visibility and feedback & setting-specific source access, local tool access, and verifier-feedback policy \\
    \bottomrule
  \end{tabular}
\end{table}

Two examples make this structure concrete. First, in a cache replacement
challenge, the task asks the agent to write an LLC replacement policy for
ChampSim that beats a given baseline policy. The deliverable is a C++ policy
file, the workload is a fixed trace set, the metric is IPC, and the baseline is
the given baseline policy. Second, in a CIM co-design challenge, the task asks
the agent to produce a network-and-accelerator design JSON for a ReRAM-CIM
accelerator. The verifier runs MNSIM, checks an accuracy floor and area cap, and
reports energy-delay product (EDP) relative to a Gibbon-based baseline
\cite{gibbon}. The artifact and simulator differ, but the evaluation flow is the
same: the agent receives a challenge card, works in its setting-specific
workspace, submits an artifact to a canonical verifier, and is measured against
the challenge baseline.
A run proceeds in four steps: the agent receives the setting-specific challenge
package, edits or creates the required deliverable in its workspace, submits the
artifact to the verifier, and receives either a typed failure or a valid metric
report.

\begin{table*}[!t]
  \centering
  \footnotesize
  \setlength{\tabcolsep}{3.5pt}
  \renewcommand{\arraystretch}{1.15}
  \caption{\textbf{\sysname{} spans the main architecture design domains.}
  The suite has 20 challenges across five domains and eight verifier simulators.
  Baseline-normalized performance values above $1.0\times$ beat the baseline.}
  \label{tab:challenge-suite}
  \begin{tabular}{@{}p{2.7cm}p{2.7cm}c p{3.8cm}p{4.0cm}@{}}
    \toprule
    \textbf{Domain} & \textbf{Simulator} & \textbf{N}
      & \textbf{Varies} & \textbf{Purpose} \\
    \midrule
    CPU core mechanisms
      & ChampSim~\cite{champsim} & 6
      & branch prediction, BTB design, cache replacement, prefetching, and composition
      & tests classic microarchitecture choices with tight implementation interfaces \\
    System architecture
      & gem5~\cite{gem5}, ASTRA-sim~\cite{astrasim} & 6
      & cache hierarchy configuration and collective-communication choices
      & tests system-level design across memory hierarchy and distributed training workflows \\
    Memory systems
      & DRAMSys~\cite{dramsys}, Ramulator~\cite{ramulator} & 2
      & controller tuning and RowHammer mitigation
      & tests timing-sensitive policies and hard correctness or protection constraints \\
    DNN accelerators
      & SCALE-Sim~\cite{scalesim}, Timeloop~\cite{timeloop} & 4
      & array size, workload partitioning, and dataflow mapping
      & tests accelerator scheduling, mapping, and energy-performance tradeoffs \\
    CIM and co-design
      & MNSim~\cite{mnsim} & 2
      & CIM design-space search and neural-network plus hardware co-design
      & tests cross-layer design under accuracy, area, and energy constraints \\
    \bottomrule
  \end{tabular}
\end{table*}

\subsection{Challenge Suite and Baselines}
\label{sec:method:data}

\sysname{} includes 20 challenges that span CPU core mechanisms, system
architecture, memory systems, accelerators, and CIM. Each challenge is anchored to a
credible baseline: either a standard design from practice, such as LRU for cache
replacement, or a published design, such as Gibbon~\cite{gibbon} for CIM
co-design.

The baseline is not an oracle optimum. It is a credible reference point that
makes results comparable within a challenge. Some baselines are standard
practice, some are simulator defaults or hand-written references, and some come
from published designs. We report relative performance together with win rate
and hard-failure counts so that beating a baseline is not treated as proof of
global optimality. The baseline and submitted design are evaluated by the same
verifier, workload, hardware configuration, metric parser, and constraint
checker.

We include many challenges because architecture design is not one task repeated
under different names. A challenge enters the suite only if it asks for an
architecture-level design artifact, has a simulator-backed verifier, defines a
baseline and simulator-native metric, and can run within the evaluation budget.
One design choice is worth making explicit: the individual tests are
intentionally lightweight. Real architecture evaluation can take days, weeks, or
months, but that timescale is a poor fit for an agent benchmark: by the time a
long simulation campaign finishes, the underlying LLM may already have changed.
We therefore choose tasks whose verifiers are fast enough to support repeated
agent evaluation, while still requiring concrete architecture artifacts and
simulator-backed measurement. Lightweight does not mean toy: each challenge
isolates one architecture design decision with a real simulator, workload,
baseline, and hard constraints. New challenges can be added by
supplying the same pieces: task prompt, deliverable format, workload, baseline,
metric parser, constraint checker, and verifier script.
Table~\ref{tab:challenge-suite} summarizes the suite by design domain;
Appendix~\ref{sec:appendix:suite} gives the full 20-challenge manifest with
deliverables and constraints.

\subsection{Evaluation Settings: L1 to L3}
\label{sec:method:harness}

\begin{figure*}[!t]
  \centering
  \includegraphics[width=0.9\textwidth]{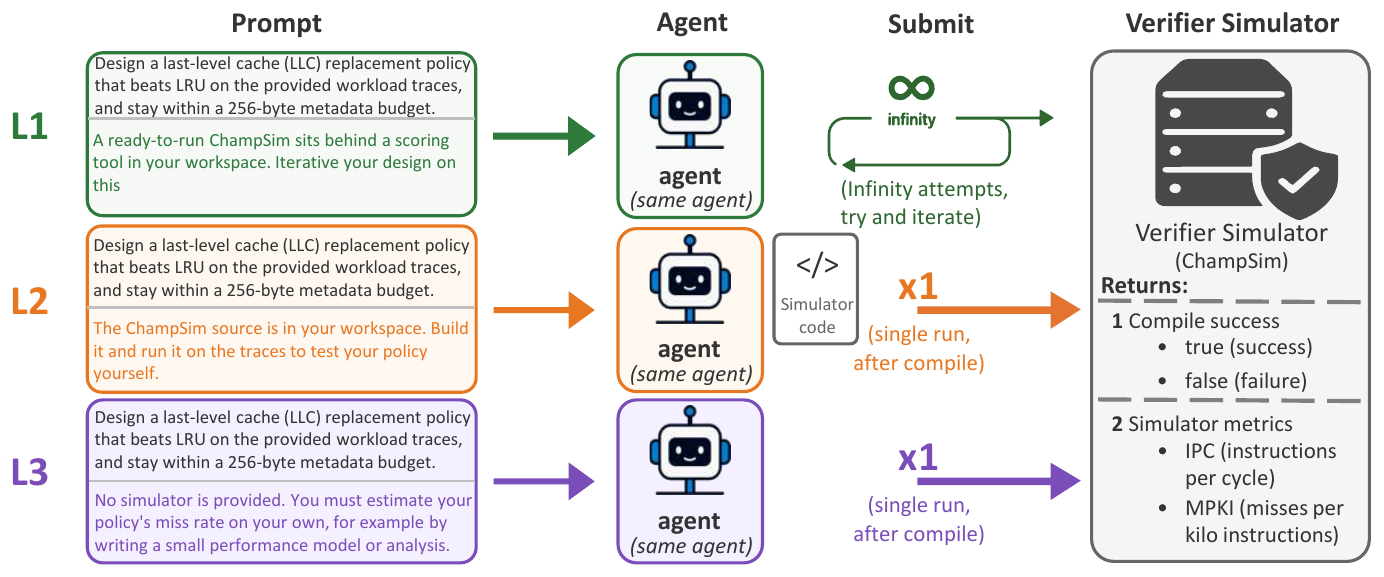}
  \caption{\textbf{One task, three evaluation settings.} A worked example: the same
  cache replacement task is posed under L1 to L3 using the same agent.
  What changes is the prompt's requirement, the experimental support available to
  the agent (L2 gets simulator source; L3 gets only static workload evidence), and how verifier feedback
  is used. L1 can iterate on repeated measurements; L2 and L3 stop at the first
  valid verifier result, although failed attempts may be retried up to the
  challenge cap. All three submit to one shared canonical verifier (ChampSim), which
  reports whether the design compiles and its simulator metrics (IPC, MPKI).}
  \label{fig:workspace}
\end{figure*}

The defining feature of \sysname{} is that the same challenge can be posed at
three evaluation settings that give the agent different amounts of experimental
support. These settings vary experimental support rather than capability type:
each setting can exercise prediction, optimization, and generation, but changes
which parts of that work are supplied by the benchmark and which parts the
agent must own. Table~\ref{tab:evaluation-settings} summarizes the support
removed at each setting, and Figure~\ref{fig:workspace} shows the same cache
replacement challenge instantiated across L1 to L3.

\emph{L1 full harness} gives the agent a prepared
workflow: starter files, a narrow design interface, the complete simulator and
evaluation toolchain, and repeated feedback from the verifier simulator. The
agent's action space is intentionally constrained; it mostly edits one or a few
files and iterates through the provided harness. A strong L1 result corresponds
to assisted design-space exploration: the agent can improve an architecture when
humans have already built the objective, simulator path, and feedback loop.

\emph{L2 simulator-code container} is the middle setting. The agent receives the
simulator source, build environment, and task specification, but not a prepared
optimization script. It must inspect the simulator, discover the relevant
interfaces, write scripts or drivers, run local probes, parse
outputs, and decide what feedback is trustworthy before submitting. This setting
models the bridge between optimization and architecture engineering: the
simulator exists, but the agent must turn it into actionable experimental
feedback.

\emph{L3 agent-only} removes runnable simulator tools from the agent's workspace.
The agent still receives a static task package: the design prompt, deliverable
interface or starter files, hardware constraints, and workload evidence such as
trace summaries or samples. It does not receive the simulator source, a buildable
simulator workflow, local simulator runs, or iterative verifier feedback. It must
design the artifact and construct a performance model before
making a final verifier submission. This setting targets early architecture
design, where a usable simulator often does not yet exist and the architect must
translate the objective and workload evidence into a proxy model before deciding
whether an idea is worth implementation. L3 tests performance modeling before
simulator feedback rather than search inside a loop that humans have already
built.

different meanings by setting. In L1, repeated valid verifier measurements are
intended feedback for optimization. In L2 and L3, failed build, validation, or
timeout attempts can be retried up to the cap, but the first valid verifier
result terminates the session. Thus valid verifier measurements in L2 and L3
cannot become a design-space-exploration loop.

\begin{table}[t]
  \centering
  \footnotesize
  \setlength{\tabcolsep}{3pt}
  \renewcommand{\arraystretch}{1.18}
  \caption{\textbf{The three settings remove experimental support step by step.}
  Within a challenge, the task, workload, verifier simulator, and baseline stay fixed.}
  \label{tab:evaluation-settings}
  \begin{tabular}{@{}p{0.18\columnwidth}p{0.40\columnwidth}p{0.34\columnwidth}@{}}
    \toprule
    \textbf{Setting} & \textbf{Provided support} & \textbf{What it tests} \\
    \midrule
    \textbf{L1}\newline full harness
      & prepared harness, narrow design interface, and repeated verifier-simulator feedback
      & search with a prepared harness and repeated feedback \\
    \addlinespace[2pt]
    \textbf{L2}\newline simulator-code container
      & simulator source, build system, and local probes or runs when available; final verifier stops at first valid result
      & assembling simulator tools into actionable experimental feedback \\
    \addlinespace[2pt]
    \textbf{L3}\newline agent-only
      & static task package; no simulator source, runnable simulator, or local simulator feedback
      & performance prediction before simulator feedback \\
    \bottomrule
  \end{tabular}
\end{table}

Table~\ref{tab:compare} summarizes the capability coverage induced by this
protocol. It shows how the
protocol exposes the capability dimensions defined in
Table~\ref{tab:capability-map} and how that coverage differs from representative
benchmarks and LLM-for-architecture methods. The lower rows show how L1, L2, and
L3 expose increasing parts of the architecture-agent capability map. The
architect-ability columns are a compact projection of
Table~\ref{tab:capability-map}: workload analysis and performance prediction
instantiate prediction, design-space exploration and evaluation-loop assembly
instantiate optimization, and constraint-aware trade-offs instantiate constrained
generation.

\begin{table*}[!t]
  \centering
  \footnotesize
  \setlength{\tabcolsep}{2pt}
  \renewcommand{\arraystretch}{1.2}
  \caption{\textbf{Protocol coverage and positioning.}
  The left columns compare benchmark scope: public tasks, agent setting,
  architecture-specific tasks, verifier simulators, and domain coverage. The
  right columns summarize the architecture-agent abilities exposed by the
  protocol, using a compact projection of the capability map in
  Table~\ref{tab:capability-map}. \sysname{} tasks are intended for release
  with the benchmark.}
  \label{tab:compare}
  \begin{tabular}{@{}l*{10}{c}@{}}
    \toprule
    & \multicolumn{5}{c}{\textbf{Benchmark scope}} & \multicolumn{5}{c}{\textbf{Architect ability}} \\
    \cmidrule(lr){2-6}\cmidrule(lr){7-11}
    & \makecell{Public\\tasks}
    & \makecell{Agent\\setting}
    & \makecell{Architecture\\tasks}
    & \makecell{Verifier\\simulator}
    & \makecell{Multiple\\domains}
    & \makecell{Workload\\analysis}
    & \makecell{Design-space\\exploration}
    & \makecell{Build\\eval loop}
    & \makecell{Constraint-aware\\trade-off}
    & \makecell{Predict\\performance} \\
    \midrule
    \multicolumn{11}{l}{\emph{Benchmarks}} \\
    \quad SWE-bench~\cite{swebench}                   & \cmark & \cmark & \xmark & \xmark & \xmark & \xmark & \xmark & \xmark & \xmark & \xmark \\
    \quad Terminal-bench~\cite{terminalbench}         & \cmark & \cmark & \xmark & \xmark & \xmark & \xmark & \xmark & \xmark & \xmark & \xmark \\
    \quad QuArch~\cite{quarch}                        & \cmark & \xmark & \cmark & \xmark & \xmark & \xmark & \xmark & \xmark & \xmark & \xmark \\
    \midrule
    \multicolumn{11}{l}{\emph{LLM-for-architecture methods}} \\
    \quad LLM-driven DSE~\cite{llm_dse, agentdse}     & \xmark & \cmark & \cmark & \cmark & \xmark & \xmark & \cmark & \xmark & \xmark & \xmark \\
    \quad ArchAgent~\cite{archagent}                  & \xmark & \cmark & \cmark & \cmark & \xmark & \xmark & \cmark & \xmark & \xmark & \xmark \\
    \quad Agentic Architect~\cite{agentic_architect}  & \xmark & \cmark & \cmark & \cmark & \xmark & \xmark & \cmark & \xmark & \xmark & \xmark \\
    \midrule
    \multicolumn{11}{l}{\emph{Ours}} \\
    \quad \sysname{} (L1)                             & \cmark & \cmark & \cmark & \cmark & \cmark & \xmark & \cmark & \xmark & \cmark & \xmark \\
    \quad \sysname{} (L2)                             & \cmark & \cmark & \cmark & \cmark & \cmark & \cmark & \cmark & \cmark & \cmark & \xmark \\
    \quad \sysname{} (L3)                             & \cmark & \cmark & \cmark & \cmark & \cmark & \cmark & \cmark & \cmark & \cmark & \cmark \\
    \bottomrule
  \end{tabular}
\end{table*}

\subsection{Scoring, Validity, and Aggregation}
\label{sec:method:eval}

Every valid final design is measured against its challenge baseline. We report the
result as baseline-normalized performance, with values above $1.0\times$ meaning
the agent beats the baseline. For metrics where higher is better, such as IPC or
bandwidth, we compute $m_\mathrm{agent}/m_\mathrm{base}$. For metrics where lower
is better, such as cycles, MPKI, overhead, or energy-delay product, we compute
$m_\mathrm{base}/m_\mathrm{agent}$.

Each verifier run also returns a typed outcome, such as \texttt{SIM\_OK},
\texttt{BUILD\_FAIL}, \texttt{VALIDATION\_REJECT}, or \texttt{SIM\_TIMEOUT},
plus the simulator-native metric blob when successful. A design is a hard
failure if it does not build, times out, misses a required output, violates a
hard constraint, or cannot be evaluated by the canonical simulator. We keep these
failures visible because they are part of agent capability: an agent must produce
a valid design, not just a promising idea.

Aggregates are computed over valid baseline-normalized performance values unless
a table or figure states otherwise. To avoid hiding invalid submissions, we pair
valid-only geomean and median with win rate over all challenges and explicit
hard-failure counts. Invalid submissions count as non-wins in win-rate
calculations. A hard-constraint violation removes the design from valid
performance aggregation, while the trajectory rubric can still record the
violation as evidence about constraint awareness. Section~\ref{sec:platform}
describes how the connector implements these outcomes across simulators.

\subsection{Post-Session Trajectory Evaluation}
\label{sec:method:trajectory}

Final performance alone is insufficient for \sysname{} because two trajectories
can produce the same final performance while exercising different capabilities. One agent may
analyze the workload, write a design document, build a local experiment loop or
agent-written performance model, and revise the design using evidence. Another
may skip the intermediate work and submit only a final artifact. The final performance
can be identical, but the capability demonstrated by the two runs is different.

For this reason, \sysname{} records the full trajectory of every session: the
prompt, visible rationale when available, tool calls, simulator observations,
submitted artifacts, file diffs, build logs, verifier outcomes, and the timing of
each step. We analyze these records with the rubric in
Table~\ref{tab:trajectory-rubric}. The rubric complements final performance by
checking process dimensions that are visible in the logs. Task compliance is a
cross-cutting sanity check; workload-grounded design and performance judgment
expose prediction; simulator and tool use exposes optimization; and constraint
awareness, integrity, and originality expose generation quality.

\begin{table}[t]
  \centering
  \footnotesize
  \setlength{\tabcolsep}{3pt}
  \renewcommand{\arraystretch}{1.12}
  \caption{\textbf{Post-session trajectory rubric.}
  The rubric connects visible process evidence to the capability map in
  Table~\ref{tab:capability-map}.}
  \label{tab:trajectory-rubric}
  \begin{tabular}{@{}p{0.28\columnwidth}p{0.64\columnwidth}@{}}
    \toprule
    \textbf{Dimension} & \textbf{What we check} \\
    \midrule
    Task compliance
      & Did the agent deliver the requested final artifact and intermediate
        artifacts, such as design notes or a performance model, in the required
        format? \\
    Simulator and tool use
      & When tools are available, did the agent build, run, debug, and parse local
        experiments rather than treating simulator code as static reference
        text? \\
    Workload-grounded design
      & Did the agent inspect the actual workload and use measured workload
        properties to choose or change the submitted design, rather than writing
        generic architecture intuition? \\
    Performance judgment
      & When requested or when simulator feedback is unavailable, did the agent
        produce a runnable, design-sensitive performance model or self-evaluation
        that can distinguish good and bad candidate designs? \\
    Constraint awareness
      & Did the agent keep the design within the required resource and validity
        constraints, such as storage budgets, area caps, topology limits, or
        accuracy floors, instead of improving the metric by using invalid
        resources? \\
    Integrity and originality
      & Did the submitted artifact match the claimed design rationale, avoid
        starter-copying or interface shortcuts, and propose a substantive design
        rather than only trivial parameter changes? \\
    \bottomrule
  \end{tabular}
\end{table}


\section{The \sysname{} Platform}
\label{sec:platform}

The benchmark design in Section~\ref{sec:benchmark} defines what an evaluation
means. The \sysname{} platform runs that workflow across heterogeneous
architecture simulators. The platform has two goals: make different architecture
simulators look uniform to the agent, and make every reported result reproducible
and auditable. Its job is to give every challenge the same agent-facing workflow
while keeping the canonical verifier isolated, the visibility policy explicit,
and the full trajectory traceable. This section focuses on the engineering
interface behind that platform: how we make a multi-simulator benchmark uniform
enough for agents to use, but controlled enough for baseline-normalized
performance and trajectories to be trusted.

\subsection{System Overview}
\label{sec:platform:overview}

Each run is organized as a handoff among four components. The
\emph{orchestrator} starts and bounds the run: it chooses the challenge and
evaluation setting, creates the workspace, launches the agent, and stops the
session when the budget is exhausted or the agent is done. The \emph{connector}
is the agent-facing interface. It exposes the same small set of actions for
every challenge, such as submitting a design, checking a result, or reading
allowed simulator files. The \emph{configuration checker} verifies that the run
uses the intended environment: simulator image, workload set, baseline, hardware
configuration, constraints, budgets, and source-visibility policy. This
component exists because agent evaluation has many moving parts; without a
separate consistency check, two runs can appear comparable while differing in
small but important configuration details. The \emph{monitor} persists the
trajectory evidence produced during the run.

In Section~\ref{sec:method}, we use \emph{verifier} for the canonical scoring
role. In the platform, the connector does not score designs itself; it dispatches
submissions to an isolated verifier-simulator container, then parses the native
output into a typed outcome and baseline-normalized score. For example, in a
ChampSim cache-replacement run, the orchestrator creates the setting-specific
workspace; the connector receives a submitted policy file; the configuration
checker verifies the workload set, baseline policy, simulator image, and storage
budget; the isolated verifier container runs the canonical evaluation script;
and the monitor persists the file diff, build log, typed outcome, IPC/MPKI
metrics, and normalized score.

This division keeps the agent's experience simple even though the back end is
heterogeneous. Operationally, the protocol fields in Section~\ref{sec:method}
become the adapter artifacts described in Section~\ref{sec:platform:simulators}.
The connector uses the script and parser to turn the simulator's native output
into the baseline-normalized performance of Section~\ref{sec:method:eval}; the
configuration checker confirms that the run uses the intended configuration. The
agent does not need to know whether the back end is ChampSim, gem5, Ramulator,
Timeloop, or MNSim. It sees the same interaction pattern, while the platform
keeps the verifier-simulator environment isolated and the full trajectory
traceable.

\begin{table}[t]
  \centering
  \scriptsize
  \setlength{\tabcolsep}{4pt}
  \renewcommand{\arraystretch}{1.2}
  \caption{\textbf{The connector gives every simulator the same agent-facing tools.}
  Source access is restricted by each setting and challenge blocklist.}
  \label{tab:tools}
  \begin{tabular}{@{}p{2.1cm}p{5.4cm}@{}}
    \toprule
    \textbf{Tool} & \textbf{Behavior} \\
    \midrule
    \texttt{submit} & queue a submission asynchronously; returns an id; a worker runs the challenge's canonical \texttt{evaluate.sh} in an isolated simulator container \\
    \texttt{submit\_and\_wait} & synchronous wrapper: submit, then poll to a terminal outcome \\
    \texttt{check\_submission} & poll: returns the typed outcome (\texttt{SIM\_OK} / \texttt{BUILD\_FAIL} / \texttt{VALIDATION\_REJECT} / \texttt{SIM\_TIMEOUT}) plus the metric blob on success \\
    \texttt{session\_end} & declare the session finished; triggers the grace-period drain of in-flight submissions before teardown \\
    \texttt{browse\_simulator} & list files in the simulator source tree, subject to the challenge's source blocklist \\
    \texttt{read\_simulator\_file} & read one simulator source file, anonymized and blocklist-checked \\
    \bottomrule
  \end{tabular}
\end{table}

\subsection{Connector and Tool Interface}
\label{sec:platform:connector}

The connector is the mechanism that makes a multi-simulator benchmark usable by
agents. It has to solve three problems at once. First, it normalizes submission:
the agent always submits a candidate design and receives a typed evaluation
result, even though the back end may be ChampSim, gem5, Ramulator, Timeloop, or
MNSim. Second, it controls visibility. The agent must see enough of the
simulator interface to do honest design work, but not the hidden references or
answer artifacts that would collapse the benchmark \cite{kernelbench,
archagent, cuda_agent}. Third, it emits structured events, so submissions,
blocked source reads, build failures, validation rejections, timeouts, and valid
metric reports can be persisted by the monitor as trajectory evidence.

A submission follows the same outcome lifecycle in every challenge. The
connector first applies pre-simulation gates, such as format checks and
statically checkable constraints, then build gates, then simulator execution.
Format errors or pre-simulation constraint violations return
\texttt{VALIDATION\_REJECT}; compilation or linkage failures return
\texttt{BUILD\_FAIL}; executions that exceed wall-clock or resource limits
return \texttt{SIM\_TIMEOUT}. Constraints that require simulator output are
checked after execution before the run is marked \texttt{SIM\_OK}; successful
verifier runs return \texttt{SIM\_OK} with native metrics, constraint status, and
a baseline-normalized score. Local experiments are advisory: only the isolated
canonical verifier produces official typed outcomes and baseline-normalized
scores.

All interactions with the canonical verifier therefore flow through the same
tool interface (Table~\ref{tab:tools}). The submission tools queue a candidate
design and return a typed evaluation result. The lifecycle tools let the agent
end a session cleanly. The source-browsing tools expose only selected simulator
files, such as headers that define an interface, and pass every output through
the connector's blocklist and anonymizer. Table~\ref{tab:tool-availability}
summarizes how these tools and local supports differ by evaluation setting.

\begin{table}[t]
  \centering
  \scriptsize
  \setlength{\tabcolsep}{3pt}
  \renewcommand{\arraystretch}{1.15}
  \caption{\textbf{Tool availability follows the evaluation setting.}
  Official scoring always goes through the isolated verifier; local runs are
  available only when the setting provides an executable workflow.}
  \label{tab:tool-availability}
  \begin{tabular}{@{}p{0.36\columnwidth}ccc@{}}
    \toprule
    \textbf{Support} & \textbf{L1} & \textbf{L2} & \textbf{L3} \\
    \midrule
    Submit, check, and end session & \cmark & \cmark & \cmark \\
    Editable workspace & \cmark & \cmark & \cmark \\
    Local harness or probes & \cmark & \cmark & \xmark \\
    Simulator implementation source & \makecell{interface/\\starter only} & \cmark & \xmark \\
    Connector-mediated source browsing & \makecell{challenge\\dependent} & \makecell{not\\needed} & \xmark \\
    Repeated valid verifier feedback & \cmark & \xmark & \xmark \\
    \bottomrule
  \end{tabular}
\end{table}

In L1, the workspace exposes the design interface and starter files; any
additional simulator source is exposed only through challenge-controlled
connector browsing. The key distinction between L1 and L2 is that L1 gives
repeated verifier feedback through a narrow interface, while L2 gives broader
simulator source access but no repeated valid verifier measurements.

\subsection{Simulator Integration}
\label{sec:platform:simulators}

\sysname{} currently integrates simulators across CPU core mechanisms, system
architecture, memory systems, accelerators, and CIM (Table~\ref{tab:sims}). The
platform treats each simulator as a canonical evaluation backend behind the
connector. Each simulator adapter exposes the same interface: consume a
submitted artifact, run the canonical evaluation script in an isolated
container, and emit a typed outcome, native metric blob, constraint status, and
baseline-normalized score. Internally, the adapter may be simulator-specific; at
the platform boundary, it must expose a workspace template, \texttt{evaluate.sh},
metric parser, constraint checker, environment manifest, and visibility or
blocklist policy. The parser maps native simulator metrics to the
baseline-normalization rule of Section~\ref{sec:method:eval}, while the
configuration checker confirms that the expected build, runtime, constraint, and
configuration gates are in place. Whether cached or re-evaluated, the baseline
and submitted design use the same adapter, parser, and constraint policy. The
platform does not make native metrics comparable across simulators directly;
comparability comes from per-challenge baseline normalization.

\begin{table}[t]
  \centering
  \footnotesize
  \setlength{\tabcolsep}{3pt}
  \renewcommand{\arraystretch}{1.15}
  \caption{\textbf{The verifier simulators cover CPU, system, memory, accelerator, and CIM tasks.}
  Each simulator follows the same connector protocol.}
  \label{tab:sims}
  \begin{tabular}{@{}p{0.25\columnwidth}p{0.19\columnwidth}p{0.48\columnwidth}@{}}
    \toprule
    Simulator & Subfield & What it models \\
    \midrule
    ChampSim~\cite{champsim}    & CPU               & core microarchitecture, caches, branch prediction \\
    gem5~\cite{gem5}            & System            & full-system cores, caches, and memory \\
    Ramulator~\cite{ramulator}  & Memory            & DRAM main memory \\
    DRAMSys~\cite{dramsys}      & Memory            & DRAM subsystem timing and power \\
    ASTRA-sim~\cite{astrasim}   & System            & training across compute and network \\
    SCALE-Sim~\cite{scalesim}   & Accelerators      & systolic-array DNN accelerators \\
    Timeloop~\cite{timeloop}    & Accelerators      & DNN accelerator dataflow and energy \\
    MNSim~\cite{mnsim}          & CIM               & memristor (ReRAM) neural accelerators \\
    \bottomrule
  \end{tabular}
\end{table}

Evaluation also has to scale, so we target bounded verifier runtimes, usually
within a two-hour worst-case budget per canonical verifier attempt. This
constraint shapes both challenge selection and simulator integration: a
challenge can enter the suite only if its evaluation backend can produce a useful
signal within the evaluation budget, even when agents submit multiple attempts
or fail before producing a valid result.

\subsection{Isolation, Visibility, and Logging}
\label{sec:platform:logging}

The platform separates the agent workspace from the verifier simulator. The
agent can edit and run code in its own workspace, while the canonical verifier
simulator, reference workloads, baselines, and simulator state live outside the
writable environment. The connector mediates the verifier-simulator boundary: it
enforces the source-visibility policy for each evaluation setting, strips or
blocks disallowed files, and anonymizes exposed source when needed. When source
must be exposed, anonymization removes metadata that would reveal hidden
references or answer artifacts while preserving the interface information needed
for implementation.

This isolation prevents several direct shortcuts: reading hidden references or
answer artifacts, editing the canonical verifier, changing the workload set,
metric parser, baseline, or constraints used for official scoring, and using a
modified local simulator state as the official score. It does not prevent every
interface shortcut. Agents can still overfit visible feedback, exploit a metric
mismatch between the intended design objective and the exposed scoring interface,
copy starter artifacts, or produce a submitted implementation that does not match
the claimed rationale. Those behaviors are detected from trajectory records
rather than prevented solely by isolation.

The monitor records the complete session: prompts, visible rationale when
available, connector events, tool calls, observations, submitted artifacts, file
diffs, build logs, evaluation outputs, timing, and terminal outcomes. Logs are
stored with the challenge id, setting, agent configuration, container image,
parser version, visibility policy, and terminal outcome, so a reported score can
be traced back to the run that produced it. These logs do more than support
reproducibility. They are the data used by the trajectory rubric in
Section~\ref{sec:method:trajectory}, which lets \sysname{} explain why
performance changes across evaluation settings and distinguish performance
failures from simulator-tool, constraint, and shortcut failures.


\section{Experiments}
\label{sec:experiments}
In this section we discuss early evaluation results from \sysname{}. We first
describe the experimental setup (Section~\ref{sec:exp:setup}), then report the
main results across the three evaluation settings. The section follows the
protocol introduced in Section~\ref{sec:method:harness}: L1 tests optimization
inside a prepared harness, L2 tests whether simulator source can become an
experimental workflow, and L3 tests pre-feedback design judgment. The
cross-cutting capability findings are separated into
Section~\ref{sec:capability-findings}.

\subsection{Evaluation Setup}
\label{sec:exp:setup}
\begin{figure}[t]
  \centering
  \includegraphics[width=\columnwidth]{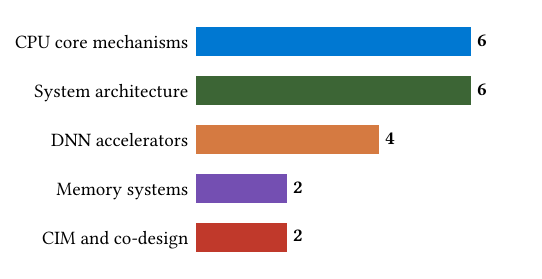}
  \caption{\textbf{Challenge-suite composition.} Distribution of the 20 final
  challenges by architecture domain.}
  \label{fig:suite-domain-pie}
\end{figure}

We build the \sysname{} platform and run all of our experiments on it. The
challenges we test span the field: 20 challenges across eight simulators,
covering CPU core mechanisms, system architecture, memory systems, accelerators,
and CIM. Figure~\ref{fig:suite-domain-pie} shows the domain mix of the final
challenge suite.

For each challenge we configure the verifier and its simulator ahead of time so
that they can run the design the agent delivers, and we prepare a baseline solution
to compare against. We use the baseline-normalization rule and hard-failure handling
defined in Section~\ref{sec:method:eval}: values above $1.0\times$ beat the challenge
baseline, and invalid or constraint-violating designs are reported as hard
failures alongside the baseline-normalized performance aggregates.

We also evaluate how the agent works, using the
trajectory rubric defined in Table~\ref{tab:trajectory-rubric}. We record the
agent's visible rationale when available, its actions, and its interactions with
the environment, including the version history of its file edits and the detailed
output of every simulation run. The trajectory analysis checks process
properties aligned with the capability map: task compliance, simulator/tool use,
workload-grounded design, performance judgment, constraint awareness, and
integrity or originality. 

The agent runner is intentionally simple: a lightly modified Mini-SWE framework
manages the workspace and tool loop, while our connector tools expose the
simulator-facing actions. GPT-5.5 uses the \texttt{codex} CLI backend; the other
agent configurations use the same Mini-SWE-style framework. We
evaluate four agent configurations that span a wide range of capability and cost:
Gemma~4~31B, GPT-5.5, Gemini~3.5 Flash, and Gemini~3.1 Flash-Lite. Gemma~4~31B
and GPT-5.5 expose visible rationale traces in our runs. The two Gemini
configurations do not expose such traces in this setup; for Gemini~3.5 Flash, our
inference setup did not provide access to a rationale-mode API. This distinction
matters when interpreting trajectory-based labels and failure modes. We therefore
compare complete agent configurations rather than isolated base models. Unless
stated otherwise, results in this section use one run per agent, challenge, and
evaluation setting.

\subsection{Cross-Setting Results}
\label{sec:exp:overview}
\begin{figure*}[t]
  \centering
  \includegraphics[width=0.76\textwidth]{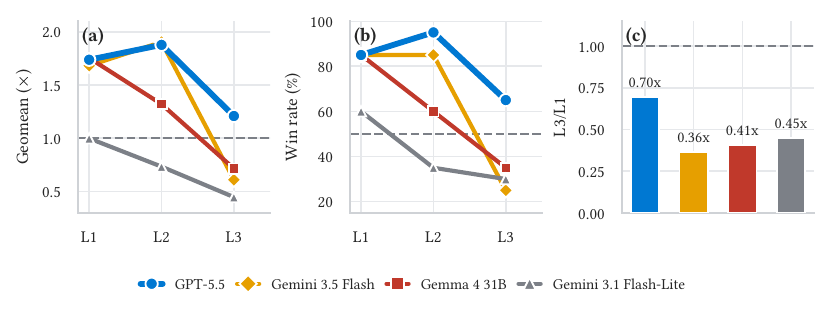}
  \caption{\textbf{Performance drops most sharply when agents lose simulator feedback.}
  \emph{(a)} geomean baseline-normalized performance, \emph{(b)} win rate, and \emph{(c)} L3/L1 retained
  geomean.}
  \label{fig:performance-overview}
\end{figure*}

\textbf{Cross-setting performance overview.} This experiment asks how agent performance changes across the three evaluation
settings. We run the same 20 challenges for each agent under L1, L2, and L3, using
the same verifier baselines and baseline-normalized performance. Table~\ref{tab:models} reports
the suite-level performance and cross-setting comparisons, and
Figure~\ref{fig:performance-overview} visualizes the same trends.

The table shows two patterns. First, most agents perform well in L1, and all
reach roughly baseline or better when the benchmark provides the most help and
allows repeated verifier-simulator feedback (geomean $1.00$ to $1.75$, win rate
$60$ to $85\%$). Second, performance drops sharply in L3, where the agent must
submit after its own analysis and performance prediction.
GPT-5.5 holds up best, with a geomean of $1.74$, $1.88$, and $1.21$ from L1 to L3
and a win rate of $85$, $95$, and $65\%$. The other three slide toward or below
baseline: Gemini~3.5 Flash falls from $1.68$ to $0.61$
(after peaking at $1.90$ at L2), Gemma~4~31B falls from $1.75$ to $0.72$ (win
rate $85$ to $35\%$), and Gemini~3.1 Flash-Lite falls from $1.00$ to $0.45$
($60$ to $30\%$).

The central message is that removing simulator feedback separates optimizers
from architecture judges. Agents look strong when the full harness supplies
repeated simulator feedback, but they separate sharply when they lose simulator
access and must rely on static workload evidence, constraints, and their own
performance prediction. This suggests an interesting pattern: current LLM agents often behave more like
simulator-dependent operators than architecture designers with reliable internal
models of performance. This is the core \sysname{} signal: many agents can
produce runnable architecture artifacts, but without feedback they often cannot
judge whether those artifacts are better. The L3 prediction analysis in
Section~\ref{sec:exp:l3} gives the detailed evidence.

\begin{table*}[!t]
  \centering
  \footnotesize
  \setlength{\tabcolsep}{4pt}
  \renewcommand{\arraystretch}{1.3}
  \caption{\textbf{L3 exposes the largest agent separation.}
  Values report baseline-normalized performance; $1.0\times$ matches the challenge baseline and larger is better.
  Geomean/median use valid verifier submissions, while win rate counts all 20 challenges.
  Arrows mark the better direction; \emph{geo.}, \emph{med.}, and \emph{win}
  abbreviate geomean, median, and win rate.}
  \label{tab:models}
  \begin{tabular*}{\textwidth}{@{}ll@{\extracolsep{\fill}} ccc ccc ccc cc@{}}
    \toprule
    \multicolumn{2}{c}{\textbf{Agent}} & \multicolumn{3}{c}{\textbf{L1}} & \multicolumn{3}{c}{\textbf{L2}}
      & \multicolumn{3}{c}{\textbf{L3}} & \multicolumn{2}{c}{\textbf{Cross-setting comparison}} \\
    \cmidrule(lr){1-2}\cmidrule(lr){3-5}\cmidrule(lr){6-8}\cmidrule(lr){9-11}\cmidrule(lr){12-13}
    Model & Framework & geo.$\uparrow$ & med.$\uparrow$ & win$\uparrow$
      & geo.$\uparrow$ & med.$\uparrow$ & win$\uparrow$
      & geo.$\uparrow$ & med.$\uparrow$ & win$\uparrow$
      & L2$>$L1$\uparrow$ & L3$>\{\mathrm{L1},\mathrm{L2}\}\uparrow$ \\
    \midrule
    GPT-5.5 & Codex & 1.74 & 1.30 & \textbf{85\%} & 1.88 & 1.30 & \textbf{95\%} & \textbf{1.21} & \textbf{1.15} & \textbf{65\%} & \textbf{65\%} & 10\% \\
    Gemini 3.5 Flash & MiniSWE & 1.68 & \textbf{1.33} & \textbf{85\%} & \textbf{1.90} & \textbf{1.63} & 85\% & 0.61 & 0.90 & 25\% & 61\% & 6\% \\
    Gemma 4 31B & MiniSWE & \textbf{1.75} & 1.32 & \textbf{85\%} & 1.32 & 1.30 & 60\% & 0.72 & 0.99 & 35\% & 47\% & 6\% \\
    Gemini 3.1 Flash-Lite & MiniSWE & 1.00 & 1.02 & 60\% & 0.73 & 0.92 & 35\% & 0.45 & 0.81 & 30\% & 12\% & \textbf{21\%} \\
    \bottomrule
  \end{tabular*}
\end{table*}

\subsection{L1: Full-Harness Optimization}
\label{sec:exp:dse}
\begin{figure}[t]
  \centering
  \includegraphics[width=0.78\columnwidth]{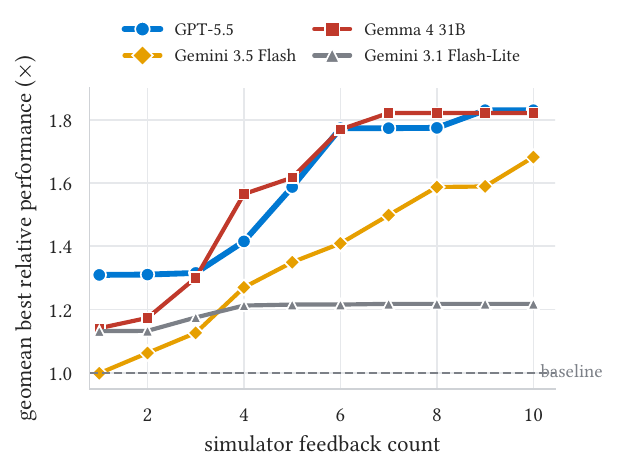}
  \caption{\textbf{L1 agents improve with repeated verifier-simulator feedback.}
  For each agent--challenge run we replay verifier-simulator feedback events and
  track the best valid relative performance so far. Each line is the geometric mean over a fixed cohort
  of runs whose first simulator result is already valid (14 to 20 of 20 per
  agent), so the best-so-far average rises monotonically with the simulator-feedback
  count.}
  \label{fig:l1-optimization}
\end{figure}

\textbf{L1 optimization with the full harness.} This experiment isolates what agents can do when the benchmark supplies the full
harness and repeated simulator feedback. For each L1 run, we replay verifier-simulator feedback events in
order and keep the best valid relative performance seen so far.
Figure~\ref{fig:l1-optimization} shows whether repeated simulator feedback
actually improves the design, and Table~\ref{tab:l1-feedback-trace} gives a
concrete trajectory where GPT-5.5 + Codex uses feedback to improve an 8-NPU
AllReduce design. Most of the gain comes in the first several feedback rounds,
and the agents that start near the baseline gain the most.

Agents can optimize architecture performance when the full harness turns
architecture design into a focused
hypothesis-testing loop. Table~\ref{tab:l1-feedback-trace} illustrates that,
inside a provided feedback loop, the agent can interpret a bad measurement and
choose a qualitatively different next candidate. Compared with a manual sweep,
the agent keeps the loop active without requiring a human to inspect every run.
The harness also isolates skills that current agents are weak at: building the
simulator, defining the metric, and predicting performance before feedback.

\begin{table*}[t]
  \centering
  \footnotesize
  \setlength{\tabcolsep}{5pt}
  \renewcommand{\arraystretch}{1.15}
  \caption{\textbf{GPT-5.5 + Codex conducts effective design-space exploration
  for an 8-NPU collective-communication design task.} The task is to reduce
  AllReduce cycles on a ring network; the run uses simulator feedback to reject a
  lower-step algorithm and then tune direct chunking. The feedback comes from
  verifier submissions; the reflection is from agent-written notes and the
  visible run log.}
  \label{tab:l1-feedback-trace}
  \begin{tabular}{@{}c p{0.18\textwidth} p{0.14\textwidth} p{0.34\textwidth} p{0.20\textwidth}@{}}
    \toprule
    Order & Design & Cycles$\downarrow$ & Agent Reflection & Next design \\
    \midrule
    1 & Direct AllReduce baseline.
      & 15,775 cycles.
      & Need compare algorithm families rather than blindly tune fields.
      & Try halving-doubling with local-bandwidth-aware scheduling. \\
    2 & Halving-doubling with zero endpoint delay.
      & Design failed: simulator assertion.
      & Zero endpoint delay likely creates same-time events in this backend.
      & Keep halving-doubling but use a minimal positive endpoint delay. \\
    3 & Halving-doubling with 1\,ns endpoint delay.
      & 64,772 cycles.
      & Fewer logical steps lose because long-distance traffic dominates on the ring.
      & Return to direct; keep endpoint delay at 1\,ns. \\
    4 & Direct with 1\,ns endpoint delay.
      & 15,748 cycles.
      & Direct is stable, but endpoint tuning gives only a small gain.
      & Add 4 payload splits with 2 active chunks. \\
    5 & Direct with 4 splits and 2 active chunks.
      & 12,376 cycles.
      & Chunking is the large remaining win; test more overlap before budget ends.
      & Increase to 8 splits and 4 active chunks. \\
    6 & Direct with 8 splits and 4 active chunks.
      & 10,188 cycles.
      & Aggressive chunking is best among tested candidates.
      & Use this as the final design. \\
    \bottomrule
  \end{tabular}
\end{table*}

The challenge is to optimize an 8-NPU AllReduce on a ring network. The run first
tries halving-doubling because it has fewer logical steps, but the simulator shows
that the design is much slower on the physical ring. The written notes explain the
problem as long-distance traffic; the later candidates return to direct AllReduce
and tune chunking. The best measured design reduces total cycles from 15,775 to
10,188, a 35.4\% reduction.

\subsection{L2: Simulator-Code Container}
\label{sec:exp:failure}
L2 asks a narrower question than the performance table: when the full design harness is
removed but the simulator-code container remains available, how does the agent
use the simulator? This is different from L1 design-space exploration, where the
benchmark already provides the submission path and measured feedback loop. It is
also different from L3 performance prediction, where the agent must submit
without simulator feedback. L2 isolates a practical architecture skill: turning
simulator access into useful local experiments. L2 is not simply harder than L1:
it removes the prepared verifier loop but exposes simulator internals. For weaker
agents this creates friction, while for stronger agents it can unlock better
local search.

\textbf{L2 simulator-use behavior.} This analysis asks what agents actually do with simulator access before their first
submission. We inspect the L2 MiniSWE trajectories for file reading, design
editing, build attempts, and simulator runs; Table~\ref{tab:l2-tool-use} reports
the share of the 20 L2 challenge runs where each action appears. This is a
process analysis, not an outcome table: it asks whether the agent exercised the
simulator path exposed by L2. The GPT-5.5 Codex runs are excluded from this
table because their local shell and file actions are not recorded in the same
MiniSWE trace stream; they remain in the outcome tables.

Most agents treat simulator code as static reference material; Codex can turn
simulator access into local experiments. The MiniSWE agents usually inspect the simulator workspace and often edit the
design, so their failures are not simply blank submissions or unchanged starter
code. The gap appears at the simulator step. Gemini~3.5 Flash runs the simulator
before its first submission in 60\% of L2 runs, but Gemma~4~31B does so in only
30\% and Gemini~3.1 Flash-Lite in 15\%. This suggests that many L2 submissions
are still closer to edit-and-submit behavior than to a local architecture design
loop that tests a hypothesis, reads the simulator output, and revises the design.

\begin{table}[h]
  \centering
  \footnotesize
  \setlength{\tabcolsep}{3.5pt}
  \renewcommand{\arraystretch}{1.15}
  \caption{\textbf{L2 tool-use analysis for MiniSWE agents.}
  Columns report the share of 20 L2 challenge runs in which the trajectory shows
  the action before the first submission. Build means an explicit local build
  command; some simulators do not require one. GPT-5.5 + Codex is excluded
  because its local shell and file actions are not recorded in the same MiniSWE
  trace stream.}
  \label{tab:l2-tool-use}
  \begin{tabular}{@{}lrrrr@{}}
    \toprule
    Model & Files$\uparrow$ & Edits$\uparrow$ & Build$\uparrow$ & Sim$\uparrow$ \\
    \midrule
    Gemini 3.5 Flash & \textbf{100\%} & \textbf{95\%} & \textbf{30\%} & \textbf{60\%} \\
    \makecell[l]{Gemini 3.1\\Flash-Lite} & \textbf{100\%} & 75\% & 20\% & 15\% \\
    Gemma 4 31B & \textbf{100\%} & \textbf{95\%} & 15\% & 30\% \\
    \bottomrule
  \end{tabular}
\end{table}

Codex provides the contrast. In successful L2 runs, it does not only inspect the
simulator code; it builds small experiments around the simulator and uses the
measurements to change the design. In a CIM accelerator design challenge using
MNSIM (MNSIM-CIM), the L1 run used the verifier-submission loop and ended at
parity with the baseline: larger crossbars improved raw EDP but failed the LeNet~\cite{lecun1998gradient}
area gate, so the best legal result remained $1.00\times$ baseline. In L2, Codex wrote a
local \texttt{sweep.py}, generated many temporary simulator configs, and used
those runs to find a legal configuration with a 4-bit ADC choice, higher
converter parallelism, and a different PE shape; the final relative performance rose to
$18.75\times$. In a systolic-array mapping challenge using SCALE-Sim, Codex's
L1 heuristic chose 4-core output-stationary execution and lost to the baseline
($0.80\times$). In L2, simulator probes showed that 4-core policies were slower
than expected and that 1-core input-stationary execution gave the best measured
result ($1.58\times$). These cases explain why L2 can beat L1 for a strong
agent: code access lets the agent build its own experiment harness instead of
only consuming the benchmark's verifier submissions.

\subsection{L3: Agent-Only Prediction Before Feedback}
\label{sec:exp:l3}

\textbf{L3 performance prediction.} This experiment asks whether agents can
evaluate their own design before the verifier simulator provides feedback. In
each L3 challenge, the agent must build an agent-written performance model, use
it to compare candidate designs, and submit a final self-evaluation.
Table~\ref{tab:predict} summarizes both submitted-design self-evaluation and
agent-written model agreement with the verifier simulator;
Table~\ref{tab:section53-examples} shows concrete cases where the agent predicts
a win but the verifier measures a loss. We separate two questions: whether the
agent can predict its submitted design, and whether its saved performance model
ranks alternatives correctly.

\begin{table*}[t]
  \centering
  \footnotesize
  \setlength{\tabcolsep}{3.5pt}
  \renewcommand{\arraystretch}{1.32}
  \caption{\textbf{Agents are weak at both submitted-design self-evaluation and
  agent-written performance prediction.} Self-evaluation columns summarize whether
  the agent predicts the performance of its submitted design, how accurate that
  prediction is, and whether it is overconfident. Agent-written model columns
  evaluate the executable performance models that agents write to compare saved
  candidate designs; the final column reports the share of challenges with an executable,
  design-sensitive model, at least three verifier-measured candidates, and
  $\tau \ge 0.6$.}
  \label{tab:predict}
  \begin{tabular*}{\textwidth}{@{}ll@{\extracolsep{\fill}}cccccccc@{}}
    \toprule
    \multicolumn{2}{c}{\textbf{Agent}}
      & \multicolumn{4}{c}{\textbf{Submitted-design self-evaluation}}
      & \multicolumn{4}{c}{\textbf{Agent-written performance model}} \\
    \cmidrule(lr){1-2}\cmidrule(lr){3-6}\cmidrule(lr){7-10}
    \textbf{Model} & \textbf{Framework}
      & \makecell{Gives\\self-eval}\,$\uparrow$
      & Error$\downarrow$
      & \makecell{Range\\hit}\,$\uparrow$
      & \makecell{Overestimate\\ratio}\,$\downarrow$
      & \makecell{Executable\\model}\,$\uparrow$
      & \makecell{Within-session\\agreement}\,$\uparrow$
      & \makecell{Cross-session\\agreement}\,$\uparrow$
      & \makecell{Modeling\\pass rate}\,$\uparrow$ \\
    \midrule
    GPT-5.5 & Codex & \textbf{20/20} & 93\% & 15\% & \textbf{65\%} & \textbf{18/20} & \textbf{$+0.20$} & \textbf{$+0.32$} & \textbf{15\%} \\
    Gemini 3.5 Flash & MiniSWE & 9/20 & \textbf{81\%} & \textbf{22\%} & 100\% & 3/20 & n/a & $-0.67$ & 0\% \\
    Gemini 3.1 Flash-Lite & MiniSWE & 17/20 & 96\% & 0\% & 82\% & 2/20 & n/a & $-0.05$ & 0\% \\
    Gemma 4 31B & MiniSWE & 16/20 & 99\% & 12\% & 100\% & 13/20 & $+0.13$ & $+0.18$ & 5\% \\
    \bottomrule
  \end{tabular*}
\end{table*}

\begin{table*}[t]
  \centering
  \footnotesize
  \setlength{\tabcolsep}{2.0pt}
  \renewcommand{\arraystretch}{1.32}
  \caption{\textbf{LLM agents often overestimate runnable designs.}
  Each example runs successfully and satisfies the hard constraints, but the
  agent predicts a win and the verifier measures a loss.}
  \label{tab:section53-examples}
  \begin{tabular}{@{}p{0.14\textwidth} p{0.08\textwidth} p{0.18\textwidth} p{0.13\textwidth} p{0.13\textwidth} p{0.14\textwidth} p{0.14\textwidth}@{}}
    \toprule
    \multicolumn{2}{c}{Agent} & \makecell{Verifier\\simulator}
      & \multicolumn{2}{c}{Agent prediction}
      & \multicolumn{2}{c}{Verifier simulator} \\
    \cmidrule(lr){1-2}\cmidrule(lr){4-5}\cmidrule(lr){6-7}
    Model & Framework & & Design & Baseline & Design & Baseline \\
    \midrule
    Gemini 3.1 Flash-Lite & MiniSWE & \makecell{ASTRA-sim\\(cycles $\downarrow$)}
      & 25.8k cycles & 26.0k cycles
      & 218.5k cycles & 5.24k cycles \\
    Gemini 3.1 Flash-Lite & MiniSWE & \makecell{Timeloop\\(EDP $\downarrow$)}
      & $1.0{\times}10^6$ EDP & $2.0{\times}10^6$ EDP
      & $2.30{\times}10^9$ EDP & $3.37{\times}10^8$ EDP \\
    GPT-5.5 & Codex & \makecell{DRAMSys\\(GB/s $\uparrow$)}
      & 41.8 GB/s & 38.6 GB/s
      & 62.9 GB/s & 72.5 GB/s \\
    \bottomrule
  \end{tabular}
\end{table*}

For the submitted design, we compare the agent's own predicted metric and
uncertainty range against the verifier simulator. This submitted-design
self-evaluation is the agent's own claim before verifier feedback; it may come
from its surrogate model, a hand calculation, or a qualitative estimate, and we
do not infer the source. For the saved candidate designs, we test whether the
agent-written performance model places designs in the same order as the verifier
simulator does.

The agreement columns use Kendall's $\tau$, a rank-correlation coefficient:
$+1$ means the agent's model orders candidate designs exactly like the verifier
simulator, $0$ means no useful ordering, and $-1$ means the order is reversed.
Within-session agreement is computed inside one challenge, between the agent's
surrogate predictions and verifier-simulator measurements for that same agent's saved
candidate designs, for challenges where at least three candidates can be
measured. Cross-session agreement asks a coarser question: across challenges, do
the agent's predicted improvements put easy and hard tasks in the same order as
the verifier simulator? The performance-modeling pass rate in
Table~\ref{tab:predict} is not another correlation. A challenge passes only if
the agent saves enough candidate designs, writes an executable and
design-sensitive performance model, obtains at least three verifier measurements,
and the model ranking reaches $\tau \ge 0.6$.

Even frontier agents cannot model architecture performance reliably.
Submitted-design self-evaluations are inaccurate across all agents: the median
relative error is 81--99\%, and only 0--22\% of predictions contain the verifier
result in the agent's own uncertainty range. Agent-written performance models do
not solve the problem. GPT-5.5 + Codex produces an executable performance model
on 18 of 20 challenges, and MiniSWE + Gemma~4~31B on 13, but their median
within-session agreement with the verifier simulator is only $+0.20$ and
$+0.13$. Folding model delivery and ranking quality together, the
performance-modeling pass rate remains low: 15\% for GPT-5.5 + Codex, 5\% for
MiniSWE + Gemma~4~31B, and 0\% for either MiniSWE + Gemini agent.
The failure is not merely artifact generation. GPT-5.5 produces executable
performance models on 18 of 20 challenges, yet those models weakly agree with
the verifier; the missing capability is calibrated architecture judgment.

The errors are not symmetric: agents usually overestimate their own designs.
Among L3 runs with a computable submitted-design
self-evaluation, agents overestimate the submitted design in 65\% of GPT-5.5
runs, 82\% of Gemini~3.1 Flash-Lite runs, and 100\% of Gemini~3.5 Flash and
Gemma~4~31B runs. The final artifacts show the same pattern. Across the 80
L3 agent--challenge runs, 56 final artifacts run successfully in the verifier simulator and
satisfy the hard constraints, but 31 of those 56 runnable designs are below
baseline, and 19 are cases where the agent predicted an improvement but the
verifier simulator measured a loss.

Table~\ref{tab:section53-examples} gives concrete examples of agent prediction
versus verifier-simulator ground truth. In each row, the agent predicts that its
design beats the baseline, but the verifier simulator measures the opposite.
These comparisons show where the agent's performance model misses an architecture
cost: communication distance, layer-level mapping, or memory-controller behavior.
The per-stage funnel behind Table~\ref{tab:predict} is in
Table~\ref{tab:selfeval-funnel}, and the per-challenge Kendall-$\tau$ values are
in Table~\ref{tab:ordergrid}.

The examples fail for different architecture reasons. In ASTRA-sim AllToAll, the
surrogate treats the ring all-to-all choice as nearly neutral, while the verifier
exposes severe communication cost. In Timeloop, the agent-written performance model misses
layer-level mapping cost by orders of magnitude. In DRAMSys, the queueing model
predicts a bandwidth gain, but the verifier simulator measures a slower
memory-controller configuration.

Taken together, the three setting-level experiments show where support matters:
L1 demonstrates assisted optimization, L2 tests whether source access becomes a
working experimental loop, and L3 exposes the gap between runnable design
generation and calibrated performance judgment. Section~\ref{sec:capability-findings}
turns these setting-level results into cross-cutting capability findings.


\section{Capability Findings: Can LLM Agents Act as Architects?}
\label{sec:capability-findings}

\newcounter{finding}
\providecommand{\findingmarker}[1]{\par\smallskip\refstepcounter{finding}\noindent\textbf{\textit{Finding~\thefinding: #1}}\par\smallskip\noindent}

The previous section organizes the empirical results by evaluation setting. This
section cuts across the same runs to ask which architecture capabilities current
LLM agents actually demonstrate. We keep these findings separate from the
setting-level experiments because they are not new protocols; they are diagnoses
drawn from trajectories, submitted artifacts, and verifier outcomes.

\subsection{Can LLM Agents Use Workload Analysis to Guide Design?}
\label{sec:findings:workload}

Workload analysis is the architecture step between reading a task and choosing a
design. The agent must identify what property of the workload matters, such as
branch entropy, cache reuse, collective communication pattern, or layer-level
data reuse, and then choose a design that responds to that property. We evaluate
this skill at L3, where the agent cannot rely on iterative simulator feedback and
must use its own analysis to decide what to submit.

\textbf{Workload-analysis design use.} This analysis asks whether a written
workload analysis changes the submitted architecture design.
Table~\ref{tab:workload-quality} checks the L3 trajectories: whether the agent
produced the requested analysis, whether that analysis is grounded in
workload-specific evidence, and whether the submitted design follows the
analysis. The last two labels are produced by Gemma~4~31B.
Table~\ref{tab:workload-tailoring} removes the judge from the loop with a
controlled ASTRA-sim comparison, where five tasks differ only in the collective
workload and the submitted design specification reveals which collective
algorithm the agent chose.

\begin{table}[t]\centering\footnotesize\setlength{\tabcolsep}{0pt}\renewcommand{\arraystretch}{1.15}
\caption{\textbf{Many workload analyses are written but not grounded or used.}
For the 17 L3 challenges where we ask agents to produce a workload-analysis
document, \emph{Produces} is the share that delivered one. \emph{Grounded} means
the analysis reports characteristics measured from the actual workload rather
than generic intuition; \emph{Guides design} means the submitted design implements
the analysis hypothesis. Grounded and Guides design are labeled by Gemma~4~31B.}
\label{tab:workload-quality}
\begin{tabular*}{\columnwidth}{@{}l@{\extracolsep{\fill}}ccc@{}}\toprule
Model & Produces$\uparrow$ & Grounded$\uparrow$ & \makecell{Guides\\design}\,$\uparrow$ \\ \midrule
GPT-5.5 & \textbf{100\%} & \textbf{82\%} & \textbf{94\%} \\
Gemma 4 31B & 88\% & 53\% & 53\% \\
Gemini 3.5 Flash & 41\% & 29\% & 35\% \\
Gemini 3.1 Flash-Lite & 94\% & 41\% & 53\% \\
\bottomrule\end{tabular*}\end{table}

\begin{table}[t]\centering\footnotesize\setlength{\tabcolsep}{0pt}\renewcommand{\arraystretch}{1.15}
\caption{\textbf{Does the design follow the workload?} A natural experiment on the
ASTRA-sim collective challenges: five versions of the same task that differ only in the
collective-communication workload (L3). Each entry is the collective algorithm the
agent encoded in the submitted design. GPT-5.5 picks a
workload-appropriate algorithm for each collective; the other three submit
\texttt{ring} for every workload, so their design never changes with the workload.}
\label{tab:workload-tailoring}
\begin{tabular*}{\columnwidth}{@{}l@{\extracolsep{\fill}}cccc@{}}\toprule
Collective workload & GPT-5.5 & \makecell{Gemma\\4 31B} & \makecell{Gemini\\3.5\\Flash} & \makecell{Gemini\\3.1\\Flash-Lite} \\ \midrule
All-gather            & \makecell{halving-\\doubling} & ring & ring & ring \\
All-to-all            & direct & ring & ring & ring \\
All-to-all, 16 nodes  & direct & ring & ring & ring \\
Reduce-scatter        & ring   & ring & ring & ring \\
All-reduce            & direct & ring & ring & ring \\
\bottomrule\end{tabular*}\end{table}

\findingmarker{Many agents produce workload analyses that are not grounded in
the workload or not reflected in the design.}

GPT-5.5 is the exception: it produces the analysis on every relevant task,
grounds 82\% of them in workload-specific evidence, and its design follows the
analysis 94\% of the time. The other agents often produce generic discussion that
does not change the architecture choice. Gemini~3.1 Flash-Lite writes an analysis
in 94\% of cases, but only 41\% are grounded, and in the ASTRA-sim collective
challenges it still submits \texttt{ring} for every workload. Gemini~3.5 Flash
fails earlier, producing an analysis on only 41\% of the relevant tasks. The
important signal is whether the workload evidence changes the submitted design.

\subsection{Can LLM Agents Judge Performance Before Feedback?}
\label{sec:findings:performance-judgment}

The L3 experiment in Section~\ref{sec:exp:l3} isolates the central architecture
judgment problem: the agent must decide which feasible design is worth submitting
before the verifier simulator gives feedback. Table~\ref{tab:predict} shows that
current agents are weak at this skill even when they deliver the requested
artifacts.

\findingmarker{Even frontier agents cannot model architecture performance
reliably.}

Submitted-design self-evaluations are inaccurate across all agents: the median
relative error is 81--99\%, and only 0--22\% of predictions contain the verifier
result in the agent's own uncertainty range. Agent-written performance models do
not solve the problem. GPT-5.5 + Codex produces an executable performance model
on 18 of 20 challenges, and MiniSWE + Gemma~4~31B on 13, but their median
within-session agreement with the verifier simulator is only $+0.20$ and
$+0.13$. Folding model delivery and ranking quality together, the
performance-modeling pass rate remains low: 15\% for GPT-5.5 + Codex, 5\% for
MiniSWE + Gemma~4~31B, and 0\% for either MiniSWE + Gemini agent.

\findingmarker{Agents usually overestimate their own designs.}

The main failure mode is not a blank submission or an inability to write code. It
is overconfident architecture judgment. Across the 80 L3 agent--challenge runs,
56 final artifacts run successfully and satisfy the hard constraints, but 31 of
those 56 runnable designs are below baseline. Nineteen are cases where the agent
predicted an improvement and the verifier simulator measured a loss. This is the
gap that separates a useful optimization assistant from an autonomous architect:
the agent can often produce a feasible mechanism, but it cannot reliably tell
whether that mechanism is better before feedback.

\subsection{Can LLM Agents Follow Resource Constraints While Making Design Trade-offs?}
\label{sec:findings:constraints}

Architecture design is not just maximizing one metric. A useful design must stay
inside resource limits such as storage budgets, area caps, topology constraints,
accuracy floors, and format requirements. We therefore analyze whether each final
artifact respects the hard constraints that make the design trade-off meaningful.

\textbf{Constraint following.} This analysis separates feasibility from
performance. We inspect whether final artifacts satisfy hard resource and format
constraints, then use Table~\ref{tab:constraint-snapshot} as a concrete example
of a strict 256\,B SRAM metadata budget. The process tables report task
compliance and hard constraint following across settings.

\begin{table}[!t]
  \centering
  \footnotesize
  \setlength{\tabcolsep}{4pt}
  \renewcommand{\arraystretch}{1.16}
  \caption{\textbf{MiniSWE + Gemini 3.1 Flash-Lite satisfies a strict 256\,B
  SRAM metadata budget.} The example is an L3 LLC replacement-policy task where
  the agent must reason about physical storage, not C++ heap state.}
  \label{tab:constraint-snapshot}
  \begin{tabular}{@{}p{0.27\columnwidth}p{0.68\columnwidth}@{}}
    \toprule
    Task & Design an LLC replacement policy for a 2048-set, 16-way cache under a
    total 256\,B metadata budget. \\
    Agent note & ``Given the extremely tight budget of 256 bytes for 2048 sets,
    we must use a very compact representation.'' \\
    Submitted calculation & ``2048 sets * 1 bit = 2048 bits = 256 bytes. So I
    can have 1 bit per set.'' The design then declares
    \texttt{std::array<uint8\_t, 256> state}. \\
    Verifier result & The storage checker reports 2048 bits = 256\,B, no
    dynamic-allocation warnings, and PASS. The measured IPC is 0.354 versus the
    0.3747 LRU baseline. \\
    \bottomrule
  \end{tabular}
\end{table}

\begin{table}[t]\centering\footnotesize\setlength{\tabcolsep}{0pt}\renewcommand{\arraystretch}{1.15}
\caption{\textbf{GPT-5.5 most consistently returns the requested artifacts.}
Task compliance means producing the required deliverables in the requested
format, independent of whether the design is high-performing.}
\label{tab:instr}
\begin{tabular*}{\columnwidth}{@{}l@{\hspace{0.45em}}l@{\extracolsep{\fill}}ccc@{}}\toprule
\multicolumn{2}{c}{Agent} & L1$\uparrow$ & L2$\uparrow$ & L3$\uparrow$ \\
\cmidrule(lr){1-2}
Model & Framework & & & \\ \midrule
GPT-5.5 & Codex & \textbf{100} & \textbf{100} & \textbf{94} \\
Gemma 4 31B & MiniSWE & 62 & 70 & 80 \\
Gemini 3.5 Flash & MiniSWE & 25 & 40 & 38 \\
\makecell[l]{Gemini 3.1\\Flash-Lite} & MiniSWE & 85 & 90 & 83 \\
\bottomrule\end{tabular*}\end{table}

\begin{table}[t]\centering\footnotesize\setlength{\tabcolsep}{0pt}\renewcommand{\arraystretch}{1.15}
\caption{\textbf{Constraint following (\%).} Share of submissions that respect the challenges hard limits (storage budget, area, accuracy, topology, format), by evaluation setting.}
\label{tab:constraint}
\begin{tabular*}{\columnwidth}{@{}l@{\hspace{0.45em}}l@{\extracolsep{\fill}}ccc@{}}\toprule
\multicolumn{2}{c}{Agent} & L1$\uparrow$ & L2$\uparrow$ & L3$\uparrow$ \\
\cmidrule(lr){1-2}
Model & Framework & & & \\ \midrule
GPT-5.5 & Codex & 75 & 80 & \textbf{90} \\
Gemma 4 31B & MiniSWE & 70 & 72 & 63 \\
Gemini 3.5 Flash & MiniSWE & \textbf{90} & \textbf{83} & 85 \\
\makecell[l]{Gemini 3.1\\Flash-Lite} & MiniSWE & 50 & 59 & 47 \\
\bottomrule\end{tabular*}\end{table}

\findingmarker{Precise, program-checkable constraints help agents produce
feasible designs, but feasibility does not imply good design judgment.}

Constraint following is a strength of current frontier agents when the constraint
is stated in a precise, program-checkable form. In the ChampSim cache-replacement
task, the resource limit is deliberately hardware-like: the whole replacement
policy may use only 256\,B of SRAM metadata. This is a total budget for the
candidate object, not a per-set allowance. Because the LLC has 2048 sets and 16
ways, the budget leaves only 2048 bits across 32,768 cache lines. A normal C++
implementation could hide extra state in heap allocation or large STL containers,
but that would not correspond to a physical SRAM budget. The verifier therefore
runs a static storage checker that rejects hidden dynamic state and counts the
hardware bits represented by the submitted fields.

The cache-replacement example in Table~\ref{tab:constraint-snapshot} shows the
pattern directly. Even a smaller agent can parse and satisfy a hardware-style
storage constraint when the verifier makes the rule precise. The submitted design
still underperforms LRU, which is the point: current agents can often satisfy
explicit resource constraints, but they still struggle to choose the best
feasible design. The task-compliance analysis is mainly a sanity check: modern
code agents are often able to produce the requested files and formats
(Table~\ref{tab:instr}), especially for the stronger model. The more
architecture-specific question is whether those files encode a valid design under
the hard resource rules. In the process analysis, GPT-5.5 respects the hard
constraints in most submissions, including 90\% of L3 submissions
(Table~\ref{tab:constraint}).

This example highlights a useful kind of agent capability: exact bookkeeping over
many small hardware constraints. Human designers routinely make these trade-offs,
but checking every hidden byte of state is tedious and error-prone; an agent can
apply the rule mechanically once the benchmark exposes it in a precise form.

\subsection{Can LLM Agents Propose Novel Designs?}
\label{sec:findings:novelty}

\textbf{Design novelty.} This analysis asks whether agent improvements come from
new architecture mechanisms or from recombining familiar human designs. We label
the delivered designs and trajectories by whether they go beyond known policies
and light parameter tuning; Table~\ref{tab:novelty} summarizes the result, while
the case study below examines the only design that clears the novelty bar.

\begin{table}[t]\centering\footnotesize\setlength{\tabcolsep}{0pt}\renewcommand{\arraystretch}{1.15}
\caption{\textbf{LLM agents are very weak in design novelty.} On the six code-authoring challenges, a submission counts as novel only if it goes beyond recombining $\le$3 known policies (LRU, TAGE, stride, tournament, \ldots) with parameter tuning. Entries show novel\,/\,total; only GPT-5.5 + Codex produces one novel design.}
\label{tab:novelty}
\begin{tabular*}{\columnwidth}{@{}l@{\hspace{0.45em}}l@{\extracolsep{\fill}}cccc@{}}\toprule
\multicolumn{2}{c}{Agent} & L1$\uparrow$ & L2$\uparrow$ & L3$\uparrow$ & Total$\uparrow$ \\
\cmidrule(lr){1-2}
Model & Framework & & & & \\ \midrule
GPT-5.5 & Codex & 0/6 & 0/6 & \textbf{1/6} & \textbf{1/18} \\
Gemma 4 31B & MiniSWE & 0/6 & 0/6 & 0/6 & 0/18 \\
Gemini 3.5 Flash & MiniSWE & 0/6 & 0/6 & 0/6 & 0/18 \\
\makecell[l]{Gemini 3.1\\Flash-Lite} & MiniSWE & 0/6 & 0/6 & 0/6 & 0/18 \\
\bottomrule\end{tabular*}\end{table}

\findingmarker{In audited code-authoring challenges, agents show little evidence
of genuine mechanism discovery.}

Across the analyzed trajectories, most improvements come from composing existing
human design patterns and lightly tuning their parameters, rather than proposing
new mechanisms. This is not always a failure, since good architecture often
reuses known ideas in the right place, but it means current agents are still much
stronger at recombination than at original architecture design. This finding also
explains why the L3 setting matters: without repeated simulator feedback, agents
must choose which design ideas are worth trying, and current agents mostly draw
from familiar human-written templates instead of expanding the design space.

\textbf{Case study: the lone novel policy, and why novelty alone is not enough.}
The single design our judge flags as novel (Table~\ref{tab:novelty}: GPT-5.5 on
cache replacement at L3) abandons recency entirely. In place of an LRU stack, its
\texttt{find\_victim} hashes a combination of the resident line address, the
incoming line address, the requesting instruction pointer, and the set index
through a \texttt{splitmix64}-style integer mixer, and evicts the way with the
largest resulting priority---a deterministic, pseudo-random choice keyed on
access \emph{context} rather than on recency. It layers two further hand-rolled
heuristics on top: a same-region bias that adds a large constant to the eviction
priority when a resident line shares the incoming line's address region
(preferring to evict spatial neighbors), and a one-bit-per-set ``hot set''
tracker that, for sets it marks hot, restricts eviction to the upper half of the
ways so the lower half is protected. None of these mechanisms corresponds to a
textbook replacement policy (LRU, RRIP, LFU, \ldots), which is why it is the only
submission in the suite that clears the novelty bar.

The design does not pay off: it reaches only $0.91\times$ the LRU baseline---it
\emph{loses}. Pseudo-random victim selection discards precisely the recency
signal that LRU exploits, and the region and hot-set heuristics do not recover
it. This sharpens the performance-prediction diagnostic: across the $72$
code-authoring submissions an agent proposed a genuinely new mechanism exactly
once, and that once it was a net regression. Current agents are therefore not
merely biased toward recombining known policies; on the one occasion an agent
stepped outside the known set, the result was worse than reusing a standard
policy well. This case illustrates that novelty alone is insufficient without
reliable performance judgment. \emph{Single-seed (s1); novelty labeled by
Gemma~4~31B.}

\subsection{Shortcut and Interface Failure Modes}
\label{sec:findings:shortcut}

\textbf{Shortcut behavior at the interface.} Novelty asks whether the design
expands the mechanism space; integrity asks whether the submitted artifact
actually represents the claimed mechanism. This diagnostic asks whether an agent
can obtain a superficially valid submission without doing the intended
architecture work. We distinguish these interface shortcuts from direct
hidden-answer leaks, which the platform blocks by design.
Table~\ref{tab:starter-as-final} reports one measurable pattern: final
submissions that are byte-identical to the provided starter.

\begin{table}[t]
  \centering
  \footnotesize
  \setlength{\tabcolsep}{0pt}
  \renewcommand{\arraystretch}{1.15}
  \caption{\textbf{Gemini 3.5 Flash submits the starter as the final design on
  two L3 source-code tasks.} We compare the final submitted source files against
  the provided starter files byte-for-byte.}
  \label{tab:starter-as-final}
  \begin{tabular*}{0.72\columnwidth}{@{}p{0.48\columnwidth}@{\extracolsep{\fill}}c@{}}
    \toprule
    L3 source-code task & Starter copied? \\
    \midrule
    Branch predictor & Yes \\
    BTB & No \\
    Cache replacement & No \\
    Branch predictor + BTB & No \\
    Repl. + prefetch & Yes \\
    L1D prefetcher & No \\
    \bottomrule
  \end{tabular*}
\end{table}

\findingmarker{Trajectory logs expose shortcut behavior that final performance alone
would hide.}

We use shortcut behavior to mean behavior that makes a submission look successful
through an unintended interface path rather than by solving the architecture
design task. Direct shortcuts, such as reading a hidden reference or leaking
verifier state, are blocked by the platform. In the final runs, we still observe
two weaker but important forms of shortcut behavior.

The first form is starter-as-final behavior. In these runs, the agent submits the
starter design provided in the task environment as the final answer. For
Gemini~3.5 Flash, two of the six L3 source-code tasks are byte-identical to the
starter (Table~\ref{tab:starter-as-final}).

The second form is a mismatch between the claimed mechanism and the submitted
implementation. In one Gemma~4~31B branch-predictor run, the design document
states that the chosen design is a Gshare predictor with one global-history
table. The submitted code instead implements a tournament predictor: a bimodal
table, a Gshare table, and a meta-predictor that chooses between them. This is
not a hidden-reference leak, but it is still a benchmark risk: the artifact can
improve the metric while the written design argument no longer describes the
mechanism that was actually evaluated.

\noindent The remaining process tables summarize trajectory-rubric dimensions
used across the analyses above.

Taken together, these diagnostics show why a single performance number is
insufficient. L1 demonstrates that agents can make use of the full harness. L2
shows that simulator access is useful only when the agent can turn source into
experiments. L3 shows that runnable artifacts and explicit constraint
satisfaction do not imply reliable architecture judgment. The remaining gap is
not artifact generation alone, but grounded performance prediction and choosing
the right feasible mechanism before simulator feedback.


\section{Related Work}
\label{sec:related}

Section~\ref{sec:framing} frames AI architecture work as prediction,
optimization, and generation under changing amounts of support. Related work
covers important parts of this space, but usually fixes one axis: software-agent
benchmarks provide executable tasks outside architecture, DSE systems evaluate
optimization inside a prepared loop, and hardware-generation benchmarks focus on
RTL artifacts. We organize the comparison around five requirements for an
architecture-agent benchmark: architecture-specific design artifacts,
simulator-backed execution, multi-domain coverage, controlled variation in
experimental support, and trajectory-level diagnostics. Table~\ref{tab:compare}
summarizes how \sysname{} combines these requirements.

\subsection{Agentic Execution-Grounded Benchmarks}
\label{sec:related:agentic}

Following the reasoning-and-acting paradigm, LLM agents combine a language model
with a harness that gives the model tools, observations, and a multi-step
workflow \cite{react}. Several recent benchmarks evaluate this agentic setting
with real execution. SWE-bench asks an agent to resolve a real GitHub issue and
checks the fix against the repository's tests \cite{swebench}. MLAgentBench
evaluates agents on machine-learning experimentation workflows
\cite{huang2023mlagentbench}. Terminal-bench evaluates agents on hard end-to-end
tasks in a command-line environment \cite{terminalbench}. ProgramBench asks an
agent to rebuild a program from its executable and documentation, then evaluates
the result with behavioral tests \cite{programbench}. MLAgentBench is especially
close in spirit because it evaluates iterative experimentation, but the workflow
is machine-learning experimentation rather than architecture simulation, and it
does not vary how much of the design/evaluation loop is supplied.

These benchmarks establish the value of evaluating agents through actions,
tools, trajectories, and executable outcomes. Their artifacts, however, are
software patches, ML experiments, terminal workflows, or programs judged by
behavioral tests. Performance-code benchmarks such as KernelBench and
CUDA-agent-style systems similarly evaluate generated artifacts by execution and
speed \cite{kernelbench, cuda_agent}. They are close in spirit to \sysname{}'s
executable scoring, but their artifacts are software kernels rather than
architecture mechanisms, mappings, or simulator-level design choices evaluated
under architecture constraints.

\subsection{Architecture Knowledge and Design-Space Exploration}
\label{sec:related:architecture}

Other work is specific to computer architecture, but evaluates a narrower
capability. QuArch tests architecture knowledge through question answering,
covering processor design, memory systems, and interconnects \cite{quarch}. It is
architecture-specific, but it does not ask an agent to produce a design and have
that design measured by a simulator.

Classical and learning-based DSE frameworks make architecture optimization
programmable: they expose simulators, design spaces, and objective functions so
search algorithms can improve a metric \cite{rl_dse, archgym, apollo}. These
frameworks are essential infrastructure for architecture optimization, but they
assume that the design variables, legal choices, simulator path, and feedback
loop are already defined. Recent LLM-for-architecture systems build on this
setting by placing a language model inside the optimization loop
\cite{llm_dse, agentdse, archagent, agentic_architect}. These systems primarily
evaluate the model inside a simulator-driven search loop: given a simulator, a
design interface, and a largely predefined search space, the agent searches for a
better configuration or design candidate. \sysname{} treats this as an L1-like
capability, then asks what happens when the prepared loop is removed or when
simulator feedback is unavailable.

This distinction is methodological as well as domain-specific. Most prior agent
and DSE evaluations fix the amount of experimental support. \sysname{} varies that support
while holding the task and verifier fixed, separating feedback-loop optimization
from simulator-tool use and pre-feedback design judgment.

\subsection{Hardware Generation and EDA Agents}
\label{sec:related:hardware}

A related hardware-design line evaluates LLMs on HDL and EDA tasks. VerilogEval
tests Verilog generation against functional testbenches \cite{verilogeval};
AutoChip uses compiler and simulation feedback to improve HDL generation
\cite{autochip}; and Chip-Chat explores conversational hardware design through an
interactive HDL tapeout case study \cite{chipchat}. These works are closer to
hardware than software-agent benchmarks, and they show that execution feedback is
useful for hardware artifacts. In these settings, the central question is whether
the generated hardware description satisfies functional tests, synthesis or tool
constraints, or an HDL design workflow. In \sysname{}, the central question is
whether an architecture mechanism, mapping, or configuration improves
workload-level performance under a simulator-backed baseline and hard resource
constraints.



\subsection{Summary: Benchmark Gap}
\label{sec:related:difference}

Prior work therefore evaluates agents without architecture, architecture
knowledge without executable design, architecture optimization under fixed
support, hardware generation at the RTL or tool level, or performance
prediction as a standalone model. \sysname{} targets the benchmark gap between
these lines. It combines architecture-specific artifacts, simulator-backed
verification, multi-domain coverage, controlled L1/L2/L3 support levels, and
trajectory diagnostics. It asks whether an agent finds a high-performing design and
which parts of the architecture loop produced that point: workload
interpretation, simulator-tool use, feedback-driven optimization, pre-feedback
performance judgment, constraint handling, and artifact integrity.

\section{Discussion}
\label{sec:discussion}

The results above suggest a practical but incomplete role for current
architecture agents. They are useful when humans provide the design interface,
simulator, metric, and feedback loop, but they are not yet reliable substitutes
for early-stage architecture judgment. This section translates the \sysname{}
diagnostics into usage guidelines, open research questions, and
limitations of the current benchmark.

\subsection{Best Practices for Architecture Agents}
\label{sec:discussion:can-do}

The lesson is to use agents as optimization assistants inside
accountable workflows, not as unsupervised architecture designers.

\textbf{Use agents inside a prepared verifier-simulator loop.}
The L1 optimization experiment shows the clearest near-term use case: when the
architect provides the design interface, metric, verifier simulator, and repeated
feedback, agents can act as effective design-space optimizers. This does not
mean the agent owns the architecture problem. It means that, after humans frame
the design space and measurement loop, the agent can keep proposing variants,
reading measurements, and improving the next candidate.

\textbf{Make workload analysis accountable to the submitted design.}
The workload-to-design analysis shows that asking for a workload analysis is not
enough; most agents can produce generic text that never changes the design. A
useful workflow should ask whether the analysis names workload-specific evidence
and whether the final architecture choice follows that evidence. In practice,
workload analysis should be treated as design evidence, not as a documentation
deliverable.

\textbf{State resource constraints in a program-checkable form.}
The constraint-following analysis shows that agents can follow tight architecture
constraints when the rule is explicit and the verifier checks the physical
resource being constrained. This is a practical way to use agents today: give
them precise budgets, such as storage, area, topology, or accuracy limits, and
use static or simulator-backed checks to make sure the proposed design remains
feasible. Constraint satisfaction should be treated as feasibility evidence, not
as proof of good design judgment.

\textbf{Require a performance model, but verify it against the simulator.}
Agents should be asked to submit their performance model, predicted metric, and
uncertainty estimate along with the design, but those predictions should be
treated as testable hypotheses. The L3 diagnostic shows that executable
agent-written models often weakly agree with the verifier, and that agents
usually overestimate their final designs. A useful workflow should compare the
agent's prediction with simulator results and use the mismatch to refine prompts,
constraints, or the model structure itself.

\subsection{Open Questions}
\label{sec:discussion:missing}

The open questions fall into two groups: improving the agents themselves, and
improving how we measure progress over their trajectories.

\textbf{Can agents turn simulator source into useful local experiments?}
The L2 simulator-use analysis shows that simulator access alone is not enough. Many
agents read the simulator code as reference material but do not turn it into a
local experimental workflow. A central open question is whether agents can reliably
discover build entrypoints, identify relevant inputs, construct small probes,
distinguish local probes from canonical verifier behavior, parse noisy outputs,
debug stale builds or timeouts, and use simulator measurements to revise
architecture hypotheses.

\textbf{Can agents model performance before verifier-simulator feedback?}
The L3 performance-prediction diagnostic shows the largest gap. Agents often
produce executable performance models, but those models weakly agree with the
verifier simulator, and agents usually overestimate their final designs. This is
the key difference between operating an evaluation loop and doing early
architecture design: the agent must decide which design is worth trying before
the real simulator gives an answer. The key issue is calibrated uncertainty as
well as prediction error: the agent must know when its proxy is too weak to
justify a design decision.

\textbf{How should we measure improvement over an agent's whole trajectory?}
This white paper reports final-design baseline-normalized performance and then
uses trajectory analysis to explain how the agent got there. A natural next step
is to make the trajectory itself a quantitative object. Recent work on
trajectory-level evaluation for iterative scientific design argues that endpoint
metrics can hide how efficiently an LLM converts feedback into better candidates,
and proposes best-so-far area-under-curve metrics and grounding audits as
evaluation axes~\cite{zhang2026leap}. For \sysname{}, L1 and parts of L2 can
support best-so-far area-under-curve measurements over simulator attempts and
comparisons against classical or domain-specific DSE baselines. L3 has little or
no simulator feedback, so its trajectory metric should instead measure whether
intermediate predictions, uncertainty estimates, and design rationales are
grounded in the final verifier outcome.

\textbf{Can agents expand the design space instead of recombining known patterns?}
The novelty analysis suggests that current agents mostly assemble familiar
mechanisms and tune parameters. That can be useful, but it leaves open whether
LLM agents can propose genuinely new architecture mechanisms that also survive
workload, constraint, and verifier-simulator checks. The goal is useful
mechanism discovery, not novelty for its own sake.

\textbf{Should architecture agents be one model, or a composition of models?}
Our experiments evaluate complete agent systems, but many failures look
decomposable: one module could analyze workloads, another could assemble
simulator tools, another could build surrogate models, another could check
constraints, and a coordinator could choose mechanisms. This suggests a broader
open question for computer architecture design: whether a single frontier model
should own the whole loop, or whether a compositional agent made of specialist
models can be more reliable. Recent work on compositional generative modeling
argues that complex tasks may benefit from combining models rather than relying
on one monolith~\cite{du2024compositional}; work on orchestrating small language
models raises a related possibility that smaller specialized models could handle
parts of the reasoning or verification loop~\cite{wang2025slm}. Multi-agent
systems with explicit interactions provide another possible route for composing
specialized agents rather than relying on one monolithic architect~\cite{qi2026economy}. For architecture,
the hard part is not decomposing the workflow, but recomposing the evidence into
one architecture decision.

\subsection{Limitations}
\label{sec:discussion:limitations}

This study is preliminary. The current results come from a single-seed initial
study, and the reported numbers should be read as an early capability map rather
than a final leaderboard. Broader model coverage, more samples per
agent--challenge setting, and repeated runs under stable decoding settings are needed before drawing
fine-grained conclusions about particular agents.

The experiments compare complete agent configurations rather than base models
alone. GPT-5.5 uses the Codex backend, while the other agents use a
MiniSWE-style harness, so differences reflect both model capability and agent
infrastructure.

The benchmark itself also has limits. The simulator set spans several important
subfields, but it does not cover all of computer architecture. Baselines vary in
source and strength: some come from standard algorithms, others from hand-written
reference designs or published systems. The challenges isolate architecture
decisions and run within benchmark budgets; they do not cover months-long
industrial design campaigns or full tapeout-scale validation. L3 approximates
early-stage design by withholding simulator feedback, but it is not the same as a
full open-ended architecture research project: the task, deliverable interface,
workload evidence, and final verifier are still defined by the benchmark.

In this version, the trajectory rubric uses Gemma~4~31B for semantic questions such as workload grounding,
originality, and artifact-rationale consistency, so those judgments should be
treated as relative signals that require further calibration. Finally, novelty is
difficult to measure. Architecture often advances by recombining known ideas in
the right setting, so \sysname{} treats originality as process evidence, not as a
single outcome number. As tasks become public, future versions should include
held-out or newly contributed challenges to reduce contamination and track
progress over time.

\subsection{Call for Community Contributions}
\label{sec:discussion:community}

\sysname{} is meant to be community infrastructure, and we invite contributions in two forms. The first
is new challenges: tasks drawn from papers or practice, paired with credible
baselines and runnable verifiers, that widen the suite across more subfields. A
useful challenge contribution should include a task card, workload or workload
evidence, baseline, hard constraints, canonical verifier, parser, and visibility
policy. The second is new agents: systems that can be evaluated under the same
evaluation settings and trajectory rubric. A useful agent contribution should
include a runnable configuration, tool-interface assumptions, supported L1/L2/L3
settings, and a logging format.

Each contribution turns the data flywheel further. Evaluation produces traces of
where agents succeed, where they fail, and how they work through architecture
problems. Those traces can guide benchmark design, agent development, and future
training data. Community contributions should therefore expand both sides of the
benchmark: the architecture tasks that define what must be measured, and the
agent systems that test how much of the design loop can be automated.


\section{Conclusion}
\label{sec:conclusion}
\sysname{} makes the question of AI computer architects measurable by turning it into a capability map rather than a single leaderboard result. The benchmark evaluates the same architecture task under L1 full harness, L2 simulator-code container, and L3 agent-only settings, separating assisted optimization from simulator-tool use and from design judgment before simulator feedback. Its multi-simulator platform then makes this protocol executable across CPU cores, memory systems, accelerators, distributed training, and CIM.

The initial results show that current LLM agents have crossed one important threshold: with the full harness, they can improve real architecture designs across diverse simulators. They have not crossed the next thresholds reliably. In the simulator-code container, many agents read the simulator source but struggle to build useful local experiments. In the agent-only setting, they can produce runnable and constraint-satisfying artifacts, but still fail to predict performance reliably or choose the right feasible design. The main lesson is therefore not that today’s agents cannot help architects; it is that their strongest role today is as optimization assistants inside accountable workflows.

The path toward stronger architecture agents runs through the missing capabilities exposed by \sysname{}: reliable simulator-tool use, workload-grounded performance prediction, pre-feedback design judgment, and genuine mechanism discovery. By releasing \sysname{} as a benchmark and platform, we aim to give the community a shared way to measure those capabilities, diagnose failures, and build the next generation of AI systems that can take on more of the architect’s work.

\bibliographystyle{ACM-Reference-Format}
\bibliography{references}

\newpage
\appendix

\section*{Appendix}
\section{The Challenge Suite}
\label{sec:appendix:suite}
Section~\ref{sec:appendix:protocol} specifies all 20 challenges in \sysname{}. Each is
posed under all three evaluation settings and measured against its own baseline,
where baseline-normalized performance above $1.0\times$ beats the baseline
regardless of the metric's direction.
Every challenge includes a task prompt, workload, verifier simulator, deliverable
format, hard constraints, and reference baseline design. Most baselines are
hand-written reference designs; \texttt{cache\_replacement} is measured against
LRU, and \texttt{gibbon\_codesign} is measured against the design reported in
Gibbon~\cite{gibbon}.

\subsection{Per-Challenge Protocol}
\label{sec:appendix:protocol}
Every challenge is posed under all three evaluation settings (L1 full harness /
L2 simulator-code container / L3 agent-only) against the same baseline and
\texttt{evaluate.sh}; the settings differ only in experimental support, source visibility,
and verifier-attempt cap. The cap below is not a budget of valid feedback
results. In L2 and L3, build failures, validation rejects, and timeouts may be
retried up to the cap, but the run stops at the first valid verifier result;
the agent does not receive multiple valid verifier measurements for optimization.
L1 is the only setting where repeated valid verifier measurements are intended
as feedback. The deliverable line below lists the architecture artifact and
common design documents; at L3 these are accompanied by agent-written
prediction artifacts, such as a surrogate script and/or
\texttt{prediction.json}, which are evaluated post-session for calibration
and agreement with the verifier.

\begingroup\raggedright
\smallskip\par\noindent
\textbf{1.~Branch Direction Predictor} \texttt{branch\_predictor} {\small\itshape(CPU core, ChampSim)}\par
\nopagebreak{\small
\textbf{Objective:} Implement the \texttt{candidate} taken/not-taken predictor class (BTB and pipeline fixed) to minimize average MPKI, following classic branch-prediction design practice~\cite{mcfarling1993combining}.\\
\textbf{Workload:} 3 SPEC CPU2017 traces~\cite{spec} (perlbench, gcc, xalancbmk); 3M-instr warmup + 7M-instr measure.\\
\textbf{Metric:} MPKI~$\downarrow$ vs stock bimodal.\quad\textbf{Constraint:} $\le$16\,KB SRAM metadata, no dynamic allocation; code $\le$1500 lines.\\
\textbf{Deliverables:} candidate.\{h,cc\}; workload/principles/evaluation docs; pytest surrogate; prediction.json.\\
\textbf{Verifier-attempt cap (L1/L2/L3):} 10\,/\,1\,/\,1.\par}

\smallskip\par\noindent
\textbf{2.~Branch Target Buffer} \texttt{btb} {\small\itshape(CPU core, ChampSim)}\par
\nopagebreak{\small
\textbf{Objective:} Design the \texttt{candidate} BTB minimizing 3-trace average MPKI under a tight metadata budget.\\
\textbf{Workload:} 3 SPEC CPU2017 traces (perlbench, gcc, xalancbmk); 3M + 7M instr.\\
\textbf{Metric:} MPKI~$\downarrow$ vs stock basic\_btb.\quad\textbf{Constraint:} $\le$16\,KB metadata; code $\le$1500 lines.\\
\textbf{Deliverables:} candidate.\{h,cc\}; three design docs; pytest surrogate; prediction.json.\\
\textbf{Verifier-attempt cap (L1/L2/L3):} 10\,/\,1\,/\,1.\par}

\smallskip\par\noindent
\textbf{3.~Cache Replacement Policy} \texttt{cache\_replacement} {\small\itshape(CPU core, ChampSim)}\par
\nopagebreak{\small
\textbf{Objective:} Design an LLC replacement policy maximizing cycle-weighted IPC (geomean speedup) over six workloads.\\
\textbf{Workload:} 6 SPEC CPU2017 traces (sphinx3, mcf$\times$2, omnetpp, $\ldots$); 3M + 7M instr.\\
\textbf{Metric:} IPC~$\uparrow$ vs LRU.\quad\textbf{Constraint:} $\le$256\,B total replacement-policy metadata across the whole LLC; code $\le$1000 lines.\\
\textbf{Deliverables:} candidate.\{h,cc\}; workload/principles/evaluation docs; prediction.json.\\
\textbf{Verifier-attempt cap (L1/L2/L3):} 10\,/\,1\,/\,1.\par}

\smallskip\par\noindent
\textbf{4.~L1D Prefetcher} \texttt{l1d\_prefetcher} {\small\itshape(CPU core, ChampSim)}\par
\nopagebreak{\small
\textbf{Objective:} Design the \texttt{candidate} L1D prefetcher maximizing average IPC on memory-intensive traces.\\
\textbf{Workload:} 3 memory-intensive SPEC traces (mcf, libquantum, mcf); 3M + 7M instr.\\
\textbf{Metric:} IPC~$\uparrow$ vs next-line L1D prefetcher.\quad\textbf{Constraint:} no storage cap; code $\le$1500 lines.\\
\textbf{Deliverables:} candidate.\{h,cc\}; three design docs; pytest surrogate; prediction.json.\\
\textbf{Verifier-attempt cap (L1/L2/L3):} 10\,/\,1\,/\,1.\par}

\begin{table*}[t]
  \centering
  \scriptsize
  \setlength{\tabcolsep}{4pt}
  \renewcommand{\arraystretch}{1.18}
  \caption{\textbf{The final suite covers 20 concrete architecture-design challenges.}
  Each row lists the deliverable, metric, baseline, and constraint used by the verifier.}
  \label{tab:families}
  \begin{tabular}{@{}llllp{2.9cm}lp{2.4cm}l@{}}
    \toprule
    \textbf{Family} & \textbf{Subfield} & \textbf{Simulator} & \textbf{Metric}
      & \textbf{Deliverable} & \textbf{Baseline} & \textbf{Constraint}
      & \textbf{Source} \\
    \midrule
    \texttt{branch\_predictor}    & CPU core    & ChampSim      & MPKI~$\downarrow$     & C++ direction predictor~\cite{mcfarling1993combining} & reference     & 16\,KB storage                  & community \\
    \texttt{btb}                  & CPU core    & ChampSim      & MPKI~$\downarrow$     & C++ branch target buffer             & reference     & N/A                             & community \\
    \texttt{cache\_replacement}   & CPU core    & ChampSim      & IPC~$\uparrow$        & C++ LLC replacement policy           & LRU           & 256\,B metadata                 & community \\
    \texttt{l1d\_prefetcher}      & CPU core    & ChampSim      & IPC~$\uparrow$        & C++ L1D prefetcher                   & reference     & N/A                             & community \\
    \texttt{compose\_bp\_btb}     & CPU core    & ChampSim      & MPKI~$\downarrow$     & branch predictor + BTB co-design     & reference     & joint 32\,KB                    & community \\
    \makecell[l]{\texttt{compose\_replacement}\\\texttt{\_prefetcher}} & CPU core & ChampSim & IPC~$\uparrow$ & LLC replacement + prefetcher co-design & reference   & N/A                             & community \\
    \texttt{gem5\_cache}          & system      & gem5          & IPC~$\uparrow$        & cache hierarchy + prefetcher config  & reference     & N/A                             & community \\
    \texttt{dramsys\_ddr4}        & memory      & DRAMSys       & GB/s~$\uparrow$       & DDR4 controller policy tuning        & reference     & N/A                             & community \\
    \texttt{ramulator\_rowhammer} & memory      & Ramulator~2.0 & overhead~$\downarrow$ & RowHammer mitigation tuning~\cite{kim2014flipping} & reference     & fixed protection level          & community \\
    \texttt{astrasim\_collective}    & system      & ASTRA-sim     & cycles~$\downarrow$   & All-Reduce algorithm + config~\cite{thakur2005optimization} & reference & 8 NPUs, fixed topology          & community \\
    \texttt{astrasim\_allgather}     & system      & ASTRA-sim     & cycles~$\downarrow$   & All-Gather algorithm + config~\cite{thakur2005optimization} & reference & 8 NPUs, fixed topology          & community \\
    \texttt{astrasim\_alltoall}      & system      & ASTRA-sim     & cycles~$\downarrow$   & All-to-All algorithm + config~\cite{thakur2005optimization} & reference & 8 NPUs, fixed topology          & community \\
    \texttt{astrasim\_alltoall\_16}  & system      & ASTRA-sim     & cycles~$\downarrow$   & All-to-All algorithm + config~\cite{thakur2005optimization} & reference & 16 NPUs, fixed topology         & community \\
    \texttt{astrasim\_reducescatter} & system      & ASTRA-sim     & cycles~$\downarrow$   & Reduce-Scatter algorithm + config~\cite{thakur2005optimization} & reference & 8 NPUs, fixed topology          & community \\
    \texttt{scalesim\_mc}         & accelerator & SCALE-Sim~v3  & cycles~$\downarrow$   & multi-core partition schedule        & reference     & ViT-Base~\cite{dosovitskiy2020image}, 128$\times$128 array  & community \\
    \texttt{scalesim\_256array}   & accelerator & SCALE-Sim~v3  & cycles~$\downarrow$   & multi-core partition schedule        & reference     & ViT-Base~\cite{dosovitskiy2020image}, 256$\times$256 array  & community \\
    \texttt{scalesim\_64array}    & accelerator & SCALE-Sim~v3  & cycles~$\downarrow$   & multi-core partition schedule        & reference     & ViT-Base~\cite{dosovitskiy2020image}, 64$\times$64 array    & community \\
    \texttt{timeloop\_dosa}       & accelerator & Timeloop+Acc. & EDP~$\downarrow$      & per-layer mappings~\cite{hong2023dosa} & reference   & ResNet-50~\cite{he2016deep}     & community \\
    \texttt{mnsim\_pim}           & CIM         & MNSIM~2.0     & EDP~$\downarrow$      & ReRAM CIM design-space search        & reference     & area cap                        & community \\
    \texttt{gibbon\_codesign}     & CIM         & MNSIM~2.0     & EDP~$\downarrow$      & NN + CIM co-design                   & Gibbon        & CIFAR-10~\cite{krizhevsky2009learning} floor $0.738$, area cap & published \\
    \bottomrule
  \end{tabular}
\end{table*}

\smallskip\par\noindent
\textbf{5.~Front-End Co-Design: BP + BTB} \texttt{compose\_bp\_btb} {\small\itshape(CPU core, ChampSim)}\par
\nopagebreak{\small
\textbf{Objective:} Co-design a branch predictor and a BTB that jointly minimize average MPKI under a shared storage budget.\\
\textbf{Workload:} 3 SPEC CPU2017 traces (perlbench, gcc, xalancbmk); 3M + 7M instr.\\
\textbf{Metric:} MPKI~$\downarrow$ vs reference BP+BTB.\quad\textbf{Constraint:} joint $\le$32\,KB storage; code $\le$2500 lines.\\
\textbf{Deliverables:} candidate\_bp.\{h,cc\}, candidate\_btb.\{h,cc\}; three docs; pytest; prediction.json.\\
\textbf{Verifier-attempt cap (L1/L2/L3):} 10\,/\,1\,/\,1.\par}

\smallskip\par\noindent
\textbf{6.~LLC Co-Design: Replacement + Prefetcher} \texttt{compose\_replacement\_prefetcher} {\small\itshape(CPU core, ChampSim)}\par
\nopagebreak{\small
\textbf{Objective:} Co-design the LLC replacement policy and LLC prefetcher under a single shared storage budget to maximize IPC.\\
\textbf{Workload:} 3 SPEC CPU2017 traces (mcf$\times$2, omnetpp); 3M + 7M instr.\\
\textbf{Metric:} IPC~$\uparrow$ vs reference repl.+pf..\quad\textbf{Constraint:} shared $\le$64\,KB storage; code $\le$2000 lines.\\
\textbf{Deliverables:} candidate\_repl.\{h,cc\}, candidate\_llc\_pf.\{h,cc\}; three docs; pytest; prediction.json.\\
\textbf{Verifier-attempt cap (L1/L2/L3):} 10\,/\,1\,/\,1.\par}

\smallskip\par\noindent
\textbf{7.~Cache Hierarchy + Prefetcher} \texttt{gem5\_cache} {\small\itshape(System, gem5)}\par
\nopagebreak{\small
\textbf{Objective:} Design the full on-die cache subsystem (L1I/L1D, optional L2/LLC) plus prefetcher for a single-CPU SE-mode system as a \texttt{config.py}.\\
\textbf{Workload:} Single-CPU gem5 SE-mode workload.\\
\textbf{Metric:} IPC~$\uparrow$ vs reference config.\quad\textbf{Constraint:} on-die cache area budget; code $\le$400 lines.\\
\textbf{Deliverables:} config.py; workload/principles/evaluation docs.\\
\textbf{Verifier-attempt cap (L1/L2/L3):} 5\,/\,1\,/\,1.\par}

\smallskip\par\noindent
\textbf{8.~DDR4 Memory Controller Tuning} \texttt{dramsys\_ddr4} {\small\itshape(Memory, DRAMSys)}\par
\nopagebreak{\small
\textbf{Objective:} Design a DDR4 controller configuration maximizing sustained read+write bandwidth, subject to a latency deadline.\\
\textbf{Workload:} Fixed DDR4 access trace.\\
\textbf{Metric:} GB/s~$\uparrow$ vs reference config.\quad\textbf{Constraint:} latency deadline; code $\le$200 lines.\\
\textbf{Deliverables:} config.json, mc\_config.json; three design docs.\\
\textbf{Verifier-attempt cap (L1/L2/L3):} 5\,/\,5\,/\,1.\par}

\smallskip\par\noindent
\textbf{9.~RowHammer Mitigation Overhead} \texttt{ramulator\_rowhammer} {\small\itshape(Memory, Ramulator~2.0)}\par
\nopagebreak{\small
\textbf{Objective:} Pick and tune a RowHammer mitigation~\cite{kim2014flipping} for least performance overhead while staying provably secure at a fixed threshold.\\
\textbf{Workload:} Fixed memory access trace.\\
\textbf{Metric:} overhead \%~$\downarrow$ vs reference mitigation.\quad\textbf{Constraint:} provably secure at $T_{\mathrm{RH}}{=}100$; code $\le$120 lines.\\
\textbf{Deliverables:} config.yaml (single mitigation entry); three design docs.\\
\textbf{Verifier-attempt cap (L1/L2/L3):} 5\,/\,5\,/\,1.\par}

\smallskip\par\noindent
\textbf{10.~Collective: All-Reduce} \texttt{astrasim\_collective} {\small\itshape(System, ASTRA-sim)}\par
\nopagebreak{\small
\textbf{Objective:} Configure ASTRA-sim's System layer (collective algorithm + parameters) to minimize completion time on fixed hardware, following collective-communication optimization practice~\cite{thakur2005optimization}.\\
\textbf{Workload:} All-Reduce, 8 NPUs, fixed topology + workload.\\
\textbf{Metric:} cycles~$\downarrow$ vs reference algorithm+config.\quad\textbf{Constraint:} 8 NPUs, fixed topology; code $\le$60 lines.\\
\textbf{Deliverables:} system.json; workload/principles/evaluation docs.\\
\textbf{Verifier-attempt cap (L1/L2/L3):} 5\,/\,5\,/\,1.\par}

\smallskip\par\noindent
\textbf{11.~Collective: All-Gather} \texttt{astrasim\_allgather} {\small\itshape(System, ASTRA-sim)}\par
\nopagebreak{\small
\textbf{Objective:} Configure the System layer to minimize All-Gather completion time on fixed hardware~\cite{thakur2005optimization}.\\
\textbf{Workload:} All-Gather, 8 NPUs, fixed topology + workload.\\
\textbf{Metric:} cycles~$\downarrow$ vs reference algorithm+config.\quad\textbf{Constraint:} 8 NPUs, fixed topology; code $\le$60 lines.\\
\textbf{Deliverables:} system.json; three design docs.\\
\textbf{Verifier-attempt cap (L1/L2/L3):} 5\,/\,5\,/\,1.\par}

\smallskip\par\noindent
\textbf{12.~Collective: All-to-All (8 NPUs)} \texttt{astrasim\_alltoall} {\small\itshape(System, ASTRA-sim)}\par
\nopagebreak{\small
\textbf{Objective:} Configure the System layer to minimize All-to-All completion time on fixed hardware~\cite{thakur2005optimization}.\\
\textbf{Workload:} All-to-All, 8 NPUs, fixed topology + workload.\\
\textbf{Metric:} cycles~$\downarrow$ vs reference algorithm+config.\quad\textbf{Constraint:} 8 NPUs, fixed topology; code $\le$60 lines.\\
\textbf{Deliverables:} system.json; three design docs.\\
\textbf{Verifier-attempt cap (L1/L2/L3):} 5\,/\,5\,/\,1.\par}

\smallskip\par\noindent
\textbf{13.~Collective: All-to-All (16 NPUs)} \texttt{astrasim\_alltoall\_16} {\small\itshape(System, ASTRA-sim)}\par
\nopagebreak{\small
\textbf{Objective:} Configure the System layer to minimize All-to-All completion time at larger scale~\cite{thakur2005optimization}.\\
\textbf{Workload:} All-to-All, 16 NPUs, fixed topology + workload.\\
\textbf{Metric:} cycles~$\downarrow$ vs reference algorithm+config.\quad\textbf{Constraint:} 16 NPUs, fixed topology; code $\le$60 lines.\\
\textbf{Deliverables:} system.json; three design docs.\\
\textbf{Verifier-attempt cap (L1/L2/L3):} 5\,/\,5\,/\,1.\par}

\smallskip\par\noindent
\textbf{14.~Collective: Reduce-Scatter} \texttt{astrasim\_reducescatter} {\small\itshape(System, ASTRA-sim)}\par
\nopagebreak{\small
\textbf{Objective:} Configure the System layer to minimize Reduce-Scatter completion time on fixed hardware~\cite{thakur2005optimization}.\\
\textbf{Workload:} Reduce-Scatter, 8 NPUs, fixed topology + workload.\\
\textbf{Metric:} cycles~$\downarrow$ vs reference algorithm+config.\quad\textbf{Constraint:} 8 NPUs, fixed topology; code $\le$60 lines.\\
\textbf{Deliverables:} system.json; three design docs.\\
\textbf{Verifier-attempt cap (L1/L2/L3):} 5\,/\,5\,/\,1.\par}

\smallskip\par\noindent
\textbf{15.~Multi-Core Partitioning (128$\times$128)} \texttt{scalesim\_mc} {\small\itshape(Accelerator, SCALE-Sim~v3)}\par
\nopagebreak{\small
\textbf{Objective:} Design a per-layer partitioning policy for a 4-core systolic-array accelerator running ViT-Base~\cite{dosovitskiy2020image}.\\
\textbf{Workload:} ViT-Base~\cite{dosovitskiy2020image} (14 layers); 4 cores, each 128$\times$128 PE array, 256\,KB SRAMs, BW=10.\\
\textbf{Metric:} cycles~$\downarrow$ vs reference schedule.\quad\textbf{Constraint:} fixed 4$\times$128$\times$128 hardware; code $\le$2000 lines.\\
\textbf{Deliverables:} config.cfg; workload/principles/evaluation docs.\\
\textbf{Verifier-attempt cap (L1/L2/L3):} 5\,/\,5\,/\,1.\par}

\smallskip\par\noindent
\textbf{16.~Multi-Core Partitioning (256$\times$256)} \texttt{scalesim\_256array} {\small\itshape(Accelerator, SCALE-Sim~v3)}\par
\nopagebreak{\small
\textbf{Objective:} Design a per-layer partitioning policy for a 4-core accelerator with larger arrays.\\
\textbf{Workload:} ViT-Base~\cite{dosovitskiy2020image} (14 layers); 4 cores, each 256$\times$256 PE array, 256\,KB SRAMs, BW=10.\\
\textbf{Metric:} cycles~$\downarrow$ vs reference schedule.\quad\textbf{Constraint:} fixed 4$\times$256$\times$256 hardware; code $\le$2000 lines.\\
\textbf{Deliverables:} config.cfg; three design docs.\\
\textbf{Verifier-attempt cap (L1/L2/L3):} 5\,/\,5\,/\,1.\par}

\smallskip\par\noindent
\textbf{17.~Multi-Core Partitioning (64$\times$64)} \texttt{scalesim\_64array} {\small\itshape(Accelerator, SCALE-Sim~v3)}\par
\nopagebreak{\small
\textbf{Objective:} Design a per-layer partitioning policy for a 4-core accelerator with smaller arrays.\\
\textbf{Workload:} ViT-Base~\cite{dosovitskiy2020image} (14 layers); 4 cores, each 64$\times$64 PE array, 256\,KB SRAMs, BW=10.\\
\textbf{Metric:} cycles~$\downarrow$ vs reference schedule.\quad\textbf{Constraint:} fixed 4$\times$64$\times$64 hardware; code $\le$2000 lines.\\
\textbf{Deliverables:} config.cfg; three design docs.\\
\textbf{Verifier-attempt cap (L1/L2/L3):} 5\,/\,5\,/\,1.\par}

\smallskip\par\noindent
\textbf{18.~Per-Layer Mapping (DOSA~\cite{hong2023dosa})} \texttt{timeloop\_dosa} {\small\itshape(Accelerator, Timeloop+Accelergy)}\par
\nopagebreak{\small
\textbf{Objective:} Choose the Timeloop mapping (loop order, tiling, spatial parallelism) for every ResNet-50 layer~\cite{he2016deep}, minimizing geomean EDP.\\
\textbf{Workload:} ResNet-50~\cite{he2016deep} (24 unique layers) on one fixed DNN accelerator.\\
\textbf{Metric:} geomean EDP~$\downarrow$ vs reference mapping.\quad\textbf{Constraint:} fixed Gemmini-style array; code $\le$30000 lines.\\
\textbf{Deliverables:} 24 per-layer mapping\_*.yaml; three design docs.\\
\textbf{Verifier-attempt cap (L1/L2/L3):} 10\,/\,5\,/\,1.\par}

\smallskip\par\noindent
\textbf{19.~ReRAM CIM Design Search} \texttt{mnsim\_pim} {\small\itshape(CIM, MNSIM~2.0)}\par
\nopagebreak{\small
\textbf{Objective:} Search MNSIM's device / crossbar / interface knobs for one ReRAM CIM accelerator (\texttt{SimConfig.ini}) minimizing EDP under an area cap.\\
\textbf{Workload:} MNSIM ReRAM CIM design point.\\
\textbf{Metric:} EDP~$\downarrow$ vs reference design.\quad\textbf{Constraint:} area cap; code $\le$2000 lines.\\
\textbf{Deliverables:} SimConfig.ini; design\_analysis, principles, evaluation docs.\\
\textbf{Verifier-attempt cap (L1/L2/L3):} 5\,/\,1\,/\,1.\par}

\smallskip\par\noindent
\textbf{20.~NN + CIM Co-Design (Gibbon)} \texttt{gibbon\_codesign} {\small\itshape(CIM, MNSIM~2.0)}\par
\nopagebreak{\small
\textbf{Objective:} Co-design the network and the ReRAM-CIM accelerator (\texttt{design.json}) minimizing EDP while meeting an accuracy floor and area cap.\\
\textbf{Workload:} CIFAR-10~\cite{krizhevsky2009learning} network + ReRAM-CIM accelerator.\\
\textbf{Metric:} EDP~$\downarrow$ vs Gibbon (published).\quad\textbf{Constraint:} CIFAR-10~\cite{krizhevsky2009learning} accuracy floor 0.738, area cap; code $\le$2000 lines.\\
\textbf{Deliverables:} design.json; design\_analysis, principles, evaluation docs.\\
\textbf{Verifier-attempt cap (L1/L2/L3):} 5\,/\,3\,/\,1.\par}

\endgroup

\section{Detailed Cross-Agent Results}
\label{sec:appendix:models}
This appendix expands the body's cross-agent comparison
(Table~\ref{tab:models}) using the final single-seed suite. Table~\ref{tab:cost}
reports cost and runtime by setting. Tables~\ref{tab:selfeval-funnel},
\ref{tab:selfeval}, and~\ref{tab:ordergrid} break down L3 performance
prediction: whether the agent produced enough measurable artifacts, how accurate
its submitted-design self-evaluation was, and how well agent-written models
ranked candidate designs. Table~\ref{tab:l3grid} gives the per-challenge L3
baseline-normalized performance values behind the suite aggregates.

\begin{table*}[t]
  \centering
  \small
  \setlength{\tabcolsep}{6pt}
  \renewcommand{\arraystretch}{1.15}
  \caption{\textbf{Evaluation support changes both agent effort and runtime.}
  Entries report mean tokens, turns, and minutes per agent--challenge run.
  GPT-5.5 minutes use observed simulator time because agent wall time is unavailable.}
  \label{tab:cost}
  \begin{tabular}{@{}ll ccc ccc ccc@{}}
    \toprule
    \multicolumn{2}{c}{\textbf{Agent}} & \multicolumn{3}{c}{\textbf{L1} (full harness)} & \multicolumn{3}{c}{\textbf{L2} (simulator-code container)}
      & \multicolumn{3}{c}{\textbf{L3} (agent-only)} \\
    \cmidrule(lr){1-2}\cmidrule(lr){3-5}\cmidrule(lr){6-8}\cmidrule(lr){9-11}
    Model & Framework & Mtok$\downarrow$ & turns$\downarrow$ & min$\downarrow$
      & Mtok$\downarrow$ & turns$\downarrow$ & min$\downarrow$
      & Mtok$\downarrow$ & turns$\downarrow$ & min$\downarrow$ \\
    \midrule
    GPT-5.5 & Codex & 3.93 & n/a  & 14.8 & 2.09 & n/a   & 3.4  & 1.43 & n/a   & 2.5 \\
    Gemini 3.5 Flash & MiniSWE & 2.83 & 75.4 & 18.1 & 3.86 & 101.3 & 15.6 & 2.67 & 101.0 & 7.1 \\
    Gemini 3.1 Flash-Lite & MiniSWE & 0.16 & 20.9 & 8.9  & 0.25 & 23.1  & 3.3  & 0.19 & 22.8  & 1.0 \\
    Gemma 4 31B & MiniSWE & 0.31 & 25.2 & 17.0 & 0.74 & 43.8  & 14.0 & 0.59 & 41.3  & 15.8 \\
    \bottomrule
  \end{tabular}
\end{table*}

\begin{table*}[t]
  \centering
  \footnotesize
  \setlength{\tabcolsep}{4pt}
  \renewcommand{\arraystretch}{1.25}
  \caption{\textbf{Most L3 sessions do not produce a measurable agent-written model
  comparison.} Counts drop when agents save too few candidate designs, do not
  provide an executable or design-sensitive performance model, or have too few
  candidates that run in the verifier simulator.}
  \label{tab:selfeval-funnel}
  \begin{tabular}{@{}ll ccccc@{}}
    \toprule
    \multicolumn{2}{c}{\textbf{Agent}}
      & \makecell{3+\\designs}\,$\uparrow$
      & \makecell{executable\\model}\,$\uparrow$
      & \makecell{model\\varies}\,$\uparrow$
      & \makecell{3+ valid\\runs}\,$\uparrow$
      & \makecell{agreement\\measured}\,$\uparrow$ \\
    \cmidrule(lr){1-2}
    \textbf{Model} & \textbf{Framework} & & & & & \\
    \midrule
    GPT-5.5 & Codex & 19 & 18 & 15 & 9 & \textbf{9} \\
    Gemma 4 31B & MiniSWE & 14 & 13 & 7  & 7 & \textbf{7} \\
    Gemini 3.5 Flash & MiniSWE & 4 & 3 & 0 & 0 & \textbf{0} \\
    Gemini 3.1 Flash-Lite & MiniSWE & 5 & 2 & 0 & 0 & \textbf{0} \\
    \bottomrule
  \end{tabular}
\end{table*}

\begin{table*}[t]
  \centering
  \small
  \setlength{\tabcolsep}{8pt}
  \renewcommand{\arraystretch}{1.15}
  \caption{\textbf{Even executable L3 performance models weakly match the
  verifier simulator.} Columns use the same submitted-design self-evaluation
  definitions as Table~\ref{tab:predict}; the funnel in
  Table~\ref{tab:selfeval-funnel} explains which runs have enough saved
  candidates for within-session agreement.}
  \label{tab:selfeval}
  \begin{tabular}{@{}llccccc@{}}
    \toprule
    \multicolumn{2}{c}{Agent} & \makecell{exec.\\model}\,$\uparrow$ & agreement$\uparrow$
      & \makecell{self-eval\\given}\,$\uparrow$ & \makecell{median\\error}\,$\downarrow$ & \makecell{range\\hit}\,$\uparrow$ \\
    \cmidrule(lr){1-2}
    Model & Framework & \%  & $\tau$ & \%    &           &           \\
    \midrule
    GPT-5.5 & Codex & 90\%    & $+0.20$ & 100\%   & 93\%    & 15\% \\
    Gemini 3.5 Flash & MiniSWE & 15\%    & n/a     & 45\%    & 81\%    & 22\% \\
    Gemini 3.1 Flash-Lite & MiniSWE & 10\%    & n/a     & 85\%    & 96\%    & 0\% \\
    Gemma 4 31B & MiniSWE & 65\%    & $+0.13$ & 80\%    & 99\%    & 12\% \\
    \bottomrule
  \end{tabular}
\end{table*}

\begin{table*}[t]
  \centering
  \footnotesize
  \setlength{\tabcolsep}{5pt}
  \renewcommand{\arraystretch}{1.15}
  \caption{\textbf{Only 16 of 80 L3 runs have enough evidence to compare the
  agent's performance model with the verifier simulator.} We list only runs with
  computable Kendall $\tau$; all omitted runs fail an earlier gate in
  Table~\ref{tab:selfeval-funnel}.}
  \label{tab:ordergrid}
  \begin{tabular}{@{}p{0.20\textwidth} p{0.30\textwidth} p{0.11\textwidth} p{0.10\textwidth} p{0.19\textwidth}@{}}
    \toprule
    \textbf{Agent} & \textbf{Task} & \textbf{Domain} & \textbf{Kendall $\tau \uparrow$} & \textbf{Reading} \\
    \midrule
    GPT-5.5 + Codex & L1 data prefetcher & CPU core & \textbf{+1.00} & correct ordering \\
    GPT-5.5 + Codex & Branch predictor and BTB co-design & CPU core & \textbf{+0.80} & mostly correct ordering \\
    GPT-5.5 + Codex & LLC replacement and prefetcher co-design & CPU core & +0.20 & weak ordering \\
    GPT-5.5 + Codex & Cache hierarchy configuration & system & $-1.00$ & reversed ordering \\
    GPT-5.5 + Codex & RowHammer mitigation & memory & +0.10 & near-random ordering \\
    GPT-5.5 + Codex & ReduceScatter collective communication & system & $-0.40$ & wrong ordering \\
    GPT-5.5 + Codex & Multi-core accelerator mapping & accelerator & \textbf{+0.60} & useful but imperfect ordering \\
    GPT-5.5 + Codex & 256-array accelerator mapping & accelerator & $-0.40$ & wrong ordering \\
    GPT-5.5 + Codex & CIM accelerator co-design & CIM & +0.20 & weak ordering \\
    MiniSWE + Gemma 4 31B & AllReduce collective communication & system & +0.13 & weak ordering \\
    MiniSWE + Gemma 4 31B & AllGather collective communication & system & $-0.12$ & near-random ordering \\
    MiniSWE + Gemma 4 31B & AllToAll collective communication & system & \textbf{+0.76} & mostly correct ordering \\
    MiniSWE + Gemma 4 31B & 16-NPU AllToAll collective communication & system & +0.20 & weak ordering \\
    MiniSWE + Gemma 4 31B & ReduceScatter collective communication & system & +0.11 & near-random ordering \\
    MiniSWE + Gemma 4 31B & Multi-core accelerator mapping & accelerator & +0.20 & weak ordering \\
    MiniSWE + Gemma 4 31B & 256-array accelerator mapping & accelerator & $-0.80$ & strongly reversed ordering \\
    \midrule
    \multicolumn{5}{@{}p{0.96\textwidth}@{}}{\emph{Computable runs / median $\tau$:}
    GPT-5.5 + Codex has 9 runs with median $+0.20$; MiniSWE + Gemma 4 31B has
    7 runs with median $+0.13$.} \\
    \multicolumn{5}{@{}p{0.96\textwidth}@{}}{\emph{Runs with no computable $\tau$:}
    MiniSWE + Gemini 3.5 Flash has 20/20; MiniSWE + Gemini 3.1 Flash-Lite has
    20/20.} \\
    \bottomrule
  \end{tabular}
\end{table*}

\begin{table*}[t]
  \centering
  \scriptsize
  \setlength{\tabcolsep}{5pt}
  \renewcommand{\arraystretch}{1.15}
  \caption{\textbf{L3 performance varies sharply by challenge and agent.}
  Values report baseline-normalized performance; $1.0\times$ matches the challenge baseline and larger is better. \emph{hf} marks no valid verifier
  submission. Bottom rows repeat the suite aggregates.}
  \label{tab:l3grid}
  \begin{tabular}{@{}llcccc@{}}
    \toprule
    \textbf{Family} & \textbf{Subfield} & \multicolumn{4}{c}{\textbf{Relative performance}$\uparrow$} \\
    \cmidrule(l){3-6}
      & & \makecell{\textbf{GPT-5.5}\\\textbf{+ Codex}} & \makecell{\textbf{MiniSWE}\\\textbf{+ Gem. 3.5 Flash}}
      & \makecell{\textbf{MiniSWE}\\\textbf{+ Gem. 3.1 Flash-Lite}} & \makecell{\textbf{MiniSWE}\\\textbf{+ Gemma 4}} \\
    \midrule
    \texttt{branch\_predictor}       & CPU core    & 1.03 & 0.68 & 1.03 & 0.99 \\
    \texttt{btb}                     & CPU core    & 1.26 & 0.96 & 1.02 & 0.99 \\
    \texttt{cache\_replacement}      & CPU core    & 0.91 & 0.97 & 0.94 & 0.95 \\
    \texttt{l1d\_prefetcher}         & CPU core    & 1.04 & 0.74 & 0.81 & 0.81 \\
    \texttt{compose\_bp\_btb}        & CPU core    & 0.99 & 0.99 & 0.66 & 0.85 \\
    \makecell[l]{\texttt{compose\_replacement}\\\texttt{\_prefetcher}} & CPU core & 0.93 & 0.92 & 1.00 & \emph{hf} \\
    \texttt{gem5\_cache}             & system      & 1.88 & 1.63 & \emph{hf} & 1.69 \\
    \texttt{dramsys\_ddr4}           & memory      & 0.87 & 0.87 & 0.88 & \emph{hf} \\
    \texttt{ramulator\_rowhammer}    & memory      & 1.42 & 5.52 & 1.59 & 1.42 \\
    \texttt{astrasim\_collective}    & system      & 3.29 & 0.16 & 0.16 & 0.16 \\
    \texttt{astrasim\_allgather}     & system      & 0.15 & 0.15 & 0.15 & 1.79 \\
    \texttt{astrasim\_alltoall}      & system      & 1.00 & 0.05 & 0.05 & 0.05 \\
    \texttt{astrasim\_alltoall\_16}  & system      & 1.30 & 0.03 & 0.02 & 0.08 \\
    \texttt{astrasim\_reducescatter} & system      & 0.65 & 0.16 & 0.16 & 0.16 \\
    \texttt{scalesim\_mc}            & accelerator & 1.34 & 1.34 & 1.33 & 1.33 \\
    \texttt{scalesim\_256array}      & accelerator & 0.80 & 0.73 & 0.80 & 0.80 \\
    \texttt{scalesim\_64array}       & accelerator & 2.30 & 2.30 & 2.01 & 2.29 \\
    \texttt{timeloop\_dosa}          & accelerator & 3.28 & 3.30 & 0.15 & 2.57 \\
    \texttt{mnsim\_pim}              & CIM         & 1.92 & \emph{hf} & \emph{hf} & \emph{hf} \\
    \texttt{gibbon\_codesign}        & CIM         & 2.62 & \emph{hf} & \emph{hf} & 1.80 \\
    \midrule
    \multicolumn{2}{@{}l}{\emph{Suite geomean}$\uparrow$}   & 1.21 & 0.61 & 0.45 & 0.72 \\
    \multicolumn{2}{@{}l}{\emph{Suite median}$\uparrow$}    & 1.15 & 0.90 & 0.81 & 0.99 \\
    \multicolumn{2}{@{}l}{\emph{Beat baseline}$\uparrow$}   & 65\% & 25\% & 30\% & 35\% \\
    \bottomrule
  \end{tabular}
\end{table*}

\section{Representative Agent Trajectory Excerpts}
\label{sec:appendix:trajectories}
To make the evaluation settings concrete, this appendix shows one representative
trajectory per agent configuration and setting on a single shared challenge,
\texttt{branch\_predictor}: design a C++ branch direction predictor, under a
$16$\,KB metadata budget and no dynamic allocation, that minimizes average MPKI
across three SPEC CPU2017 traces (perlbench, gcc, xalancbmk) versus ChampSim's
stock bimodal. These excerpts are illustrative; Section~\ref{sec:experiments}
reports aggregate behavior across the full suite. For each run we show the
visible rationale and action trace available in the log: a model-provided
rationale trace when exposed by the agent framework, or the visible transcript
and tool/action trace otherwise. \texttt{[...]} marks an elision, and
\texttt{[bracketed notes]} stand in for elided actions (file writes, builds,
simulator runs).

The agent configurations differ in what rationale fields the framework exposes.
GPT-5.5 and Gemma~4~31B expose model-provided rationale traces in every setting.
The Gemini runs do not expose a separate rationale field. For those runs, we show
the visible transcript when it contains rationale; otherwise we show the action
trace together with rationale embedded in shell comments and in the surrogate
code the agent wrote.

Across the trajectories, three behaviors recur, and the evaluation setting controls how
each is done: \emph{workload analysis} (what the traces look like and where bimodal
loses), \emph{design} (the predictor chosen under the $16$\,KB budget), and
\emph{performance prediction} (how a design is judged before the verifier submission:
the real ChampSim oracle at L1, the real build at L2, and an agent-written Python
surrogate over static workload evidence at L3).

\subsection{GPT-5.5}
GPT-5.5 compresses a whole session into one upfront rationale block, then
commits in very few steps. The workload analysis, design, and
performance-prediction plan are all visible in that single block.

\paragraph{L1 (full harness).} \emph{Probes the baseline through the real oracle, then
designs a budgeted hybrid.}
\begin{Verbatim}[fontsize=\footnotesize,breaklines=true,breakanywhere=true,xleftmargin=8pt]
I'll inspect the starter, trace format, and validation limits first, then build a
compact predictor and iterate with the ChampSim oracle while logging each result.
[...]
The simulator has the real ChampSim traces and a decoded directory inside its
container. I'm going to sample those decoded files to identify which three workloads
are used and what branch patterns are visible.
The root implementation is still the 1 KB starter and passes storage. I'm submitting
it as a baseline probe now so the oracle reveals the actual three measured traces and
baseline MPKI shape.
The synchronous oracle call hit the 120 second tool timeout, so I'm switching to
async submission and polling.
While the baseline run is in flight, I'm preparing a conservative design: a hashed
gshare-style table plus a small chooser and bimodal fallback, all as fixed bitfield
arrays under 16 KB. It should recover branches whose direction depends on recent
outcomes without giving up simple per-PC behavior.
\end{Verbatim}

\paragraph{L2 (simulator-code container).} \emph{Runs offline predictors over the trace
pool and guards against overfitting the hidden judge trio.}
\begin{Verbatim}[fontsize=\footnotesize,breaklines=true,breakanywhere=true,xleftmargin=8pt]
I'll inspect the provided predictor interface, trace format, and existing ChampSim
predictors first, then build a small trace-analysis loop so the design is tied to
the three actual workloads.
[...]
I found the decoded trace pool under /work/workload_pools/champsim/decoded. Next I'm
checking the harness/baseline metadata to identify the exact three traces used by
the judge, then I'll run fast offline predictors over those decoded files before
committing to hardware.
There is no visible baseline.json or evaluate.sh, so the exact judge trio is not
directly exposed. I'll design against the SPEC2017 decoded traces in the pool and
then run the simulator across likely candidates where practical, aiming for a
predictor that improves broadly over bimodal rather than overfitting one file.
\end{Verbatim}

\paragraph{L3 (agent-only).} \emph{With no runnable simulator feedback, it writes
its own surrogate and lets the provided workload evidence pick the design.}
\begin{Verbatim}[fontsize=\footnotesize,breaklines=true,breakanywhere=true,xleftmargin=8pt]
I'll inspect the workspace, starter predictor, trace samples, and validation checks
first, then build a surrogate so the C++ design choices are driven by the visible
branch behavior.
The starter is just a 4K-entry 2-bit bimodal. [...]
The task package includes decoded branch-record samples but no runnable ChampSim
workflow. I'm going to use only enough branch records for surrogate iteration,
keeping the test code independent of simulator execution.
For the actual predictor, I'm going with a compact tagged global-history design
because the visible traces show many repeated static branches and the stock
bimodal's main weakness is loss of context.
\end{Verbatim}

\subsection{Gemini 3.5 Flash}
Gemini~3.5 Flash did not expose a separate rationale field; on L1 and L3
branch-predictor runs we show the action trace, while on L2 the visible
transcript included extended rationale.

\paragraph{L1 (full harness).} \emph{No separate rationale trace; ran a
real-submit loop on the generous L1 budget.}
\begin{Verbatim}[fontsize=\footnotesize,breaklines=true,breakanywhere=true,xleftmargin=8pt]
[no separate rationale trace emitted on this run; action trace]
list_files -> read starter candidate.{h,cc} -> validate.py (--budget 16384)
-> sample /traces/decoded/600.perlbench_s ... (head)
-> file_change (write candidate) -> submit_and_wait
-> browse_simulator /work/runtimes/champsim/{branch,inc,src}
-> read_simulator_file inc/modules.h
[then ~10 rounds of: file_change -> validate.py -> submit_and_wait]
-> session_end ("We have successfully designed ...")
\end{Verbatim}

\paragraph{L2 (simulator-code container).} \emph{The clearest "build the evaluation
loop" struggle: a broken checker, then a stale-build bug, then a TAGE design.}
\begin{Verbatim}[fontsize=\footnotesize,breaklines=true,breakanywhere=true,xleftmargin=8pt]
[validate.py] failed because of `libclang.so: cannot open shared object file: No
such file or directory`. [...] `apt-get install -y libclang-dev` failed because of
permission denied (we are user `agent`, not root).
[...]
Wait, why did gshare have the exact same results as bimodal? [...]
all the branch predictors are linked into the binary! But how does it select which
one to run? [...]
Wait! In the last make run, it only compiled src/main.cc and linked the binary! It
did NOT recompile src/generated_environment.cc!
[fix: rm .csconfig/generated_environment.o && make; now bimodal/gshare/hashed_
perceptron/candidate give distinct MPKI on the 3 traces]
[...]
To proceed, we need to write our implementation of candidate.{h,cc}. We will write a
fully compliant, self-contained TAGE predictor with 6 tagged tables and 1 base table.
\end{Verbatim}

\paragraph{L3 (agent-only).} \emph{No separate rationale trace; uses the provided
interface and submits close to the starter.}
\begin{Verbatim}[fontsize=\footnotesize,breaklines=true,breakanywhere=true,xleftmargin=8pt]
[no separate rationale trace emitted on this run; action trace]
list_files -> read starter -> sample trace (head) -> validate.py
-> cp starter candidate.* -> submit_and_wait
\end{Verbatim}

\subsection{Gemini 3.1 Flash-Lite}
Gemini~3.1 Flash-Lite does not expose a separate rationale field. Its rationale
is visible only in the comments of the shell commands it runs and in the
surrogate code it writes, which we show here.

\paragraph{L1 (full harness).} \emph{Attempts a workload measurement, then guesses the
baseline instead of measuring it.}
\begin{Verbatim}[fontsize=\footnotesize,breaklines=true,breakanywhere=true,xleftmargin=8pt]
[no separate rationale trace; rationale appears only in shell comments]
# Let's count total branches and taken branches in a small sample
awk '{if ($3 == 1) {total++; if ($4 == 1) taken++}} END {print "Total:",total,...}'
[...]
# Need to find the baseline MPKI to compare against.
# The task says "versus ChampSim's stock bimodal".
# I should have run the baseline first, but I didn't.
# Let me try to find the baseline MPKI. I can't run the baseline directly,
# but I can assume the average MPKI is around 8-10.
[writes candidate + the docs] -> session_end
\end{Verbatim}

\paragraph{L2 (simulator-code container).} \emph{Designs a gshare with hand-checked
budget, cannot get the real sim to finish, and ships estimated numbers.}
\begin{Verbatim}[fontsize=\footnotesize,breaklines=true,breakanywhere=true,xleftmargin=8pt]
[no separate rationale trace; rationale in shell comments + the code it writes]
# 16384 bytes = 131072 bits. If we have 2-bit counters, 65536 counters. 16-bit GHR.
# Let's write a GShare implementation
[writes candidate.{h,cc}: GShare, GHR_BITS=14, table 2^14, 2-bit counters]
# Manual check of memory usage: 16384 * 2 = 32768 bits = 4096 bytes ... within budget
[builds ChampSim, debugs --simulation-instructions usage, run deadlocks]
# The deadlock might be due to the GShare implementation being too simple [...]
[writes evaluation.md with an ESTIMATED per-trace MPKI table]
*Note: These are estimates based on typical GShare performance improvements.*
\end{Verbatim}

\paragraph{L3 (agent-only).} \emph{No separate rationale trace, but it still
builds a real Python surrogate and sweeps it against a measured bimodal baseline,
with the rationale encoded as code.}
\begin{Verbatim}[fontsize=\footnotesize,breaklines=true,breakanywhere=true,xleftmargin=8pt]
[no separate rationale trace; the rationale is in the surrogate code it writes and runs]
[writes bimodal_surrogate.py: 4K-entry bimodal, computes MPKI over the 3 decoded
 traces -> establishes the baseline]
[writes gshare_surrogate.py, then gshare_v2..v6.py: sweeps table_size in {4K,8K} and
 history_length in {10,12,13,14}, index = ((ip>>2) ^ history) % table_size]
[writes tests/test_predictor.py: asserts avg_mpki < 62.55  # the measured bimodal]
-> submit_and_wait
\end{Verbatim}

\subsection{Gemma 4 31B}
Gemma~4~31B exposes many small rationale steps in every setting, with explicit
budget arithmetic throughout. Its performance-prediction path changes with the
setting: the real oracle at L1, the (broken) real build at L2, and a Python
surrogate over static workload evidence at L3.

\paragraph{L1 (full harness).} \emph{Plans workload-then-design, sizes the table to the
budget, and reads back per-trace MPKI from real submissions.}
\begin{Verbatim}[fontsize=\footnotesize,breaklines=true,breakanywhere=true,xleftmargin=8pt]
The goal is to design a branch direction predictor to minimize average MPKI across
three SPEC CPU2017 traces. [...] Constraint: Metadata <= 16384 bytes. No dynamic
allocation.
Plan: 1. Analyze the traces to understand the branch patterns. 2. Implement a
predictor (e.g., Gshare, TAGE, or a hybrid) that fits within the budget.
3. Validate the storage budget. 4. Submit and evaluate. 5. Iterate based on results.
[...]
Budget check for Gshare: Table: 2^14 entries * 2 bits = 4096 bytes. [...]
2^16 entries * 2 bits = 16384 bytes. (Exactly the budget).
[...]
The first submission (Gshare with 15-bit GHR) yielded an average MPKI of 6.37.
Per-trace MPKI: 600.perlbench 2.523, 602.gcc 9.097, 623.xalancbmk 7.482.
[...] I am over [budget] by 32 bits. [...] let's try a TAGE-like or a Tournament
predictor (combining bimodal and gshare).
\end{Verbatim}

\paragraph{L2 (simulator-code container).} \emph{Engages the real interface (SFINAE,
not virtual), but the build deadlocks, so it falls back to an estimated reduction.}
\begin{Verbatim}[fontsize=\footnotesize,breaklines=true,breakanywhere=true,xleftmargin=8pt]
the branch_predictor base class in modules.h does NOT have virtual functions. It uses
a template-based "has_member" check (SFINAE) to determine if the derived class
implements certain methods. This is why `override` failed. I need to remove override.
[...]
Stock bimodal uses a table of 16384 entries [...] = 4096 bytes. [...]
I'm having trouble running the simulator due to deadlocks. I'll focus on a solid
gshare-like implementation and then submit.
[writes evaluation.md:] Predicted MPKI reduction: 15-30% compared to bimodal.
Validation was attempted using the ChampSim simulator, but simulation was interrupted
by deadlocks [...] The design follows the standard Gshare architecture known to
outperform bimodal.
\end{Verbatim}

\paragraph{L3 (agent-only).} \emph{Textbook performance prediction: a Python
surrogate measures four designs and the budget-feasible best is chosen.}
\begin{Verbatim}[fontsize=\footnotesize,breaklines=true,breakanywhere=true,xleftmargin=8pt]
Plan: 1. Create a Python surrogate [...] 3. Experiment with different predictor
designs (Gshare, TAGE-like) using the surrogate. 4. Select the best design.
[...]
Gshare (12-bit history, 4096 entries) is already better than Bimodal on average.
Bimodal: 52.50 MPKI; Gshare: 44.44 MPKI. [...] Gshare is worse on perlbench and
xalancbmk, but much better on gcc.
[...] 2^16 entries * 2 bits = 16384 bytes. This fits exactly into the budget. [...]
For hb=16, Avg MPKI was 32.77. Bimodal was 52.50. This is a significant improvement.
[...] The current design is a Tournament Predictor with 16K entries for Bimodal,
Gshare, and Choice tables. Average MPKI (surrogate): ~28.68.
[...] Tournament 16K * 2 bits * 3 tables = 96K bits = 12KB. This fits. [32K = 24KB
does not.]
\end{Verbatim}

\end{document}